\documentclass[11pt]{article}

\usepackage[utf8]{inputenc}
\usepackage{graphicx}
\usepackage[colorlinks,allcolors=cyan]{hyperref}
\usepackage{amsmath,amssymb,amsfonts,mathtools}

\usepackage{lipsum}
\usepackage{wrapfig}
\usepackage{graphicx,overpic}
\usepackage{siunitx}
\usepackage{textcomp}
\usepackage{titlesec}

\usepackage{subcaption}
\usepackage{graphicx}
\graphicspath{{img/}} % includegraphics path
\newcommand{\includegraphicsw}[2][1.]{\includegraphics[width=#1\linewidth]{#2}}
\usepackage{multirow}
\usepackage{hhline}

\usepackage{bm}
\newcommand{\vect}[1]{\boldsymbol{\mathbf{#1}}}

\newcommand*\diff{\mathop{}\!\mathrm{d}}

\begin{document}

\title{Experimental validation of a phase-field model to predict coarsening dynamics of lipid domains in multicomponent membranes}

\author{A.~Zhiliakov$^1$, Y.~Wang$^2$, A.~Quaini$^{1}$, M.~Olshanskii$^{1}$, S.~Majd$^2$}

\maketitle

\begin{center}
\noindent $^{1}$Department of Mathematics, University of Houston, 3551 Cullen Blvd, Houston TX 77204\\
\texttt{alex@math.uh.edu; aquaini@central.uh.edu; molshan@math.uh.edu}

\vskip .3cm
\noindent $^{2}$Department of Biomedical Engineering, University of Houston, 3551 Cullen Blvd, Houston TX 77204\\
\texttt{ywang147@uh.edu; smajd9@central.uh.edu}
\end{center}

\vskip .3cm
\noindent{\bf Abstract}
Membrane phase-separation is a mechanism that biological membranes often use to
locally concentrate specific lipid species in order to organize diverse membrane processes.
Phase separation has also been explored as a tool for the design of liposomes
with heterogeneous and spatially organized surfaces. These ``patchy'' liposomes are promising platforms for delivery purposes,
however their design and optimization through experimentation can be expensive and time-consuming.
We developed a computationally efficient method based on the surface Cahn--Hilliard phase-field model to complement
experimental investigations in the design of patchy liposomes. The method
relies on thermodynamic considerations to set the initial state for numerical simulations.
We show that our computational approach delivers not only qualitative pictures,
but also accurate \emph{quantitative} information about the dynamics of the membrane organization.
In particular, the computational and experimental results are in excellent agreement in terms of raft area fraction,
total raft perimeter over time and total number of rafts over time for two different membrane compositions
(DOPC:DPPC with a 2:1 molar ratio with 20\% Chol and DOPC:DPPC with a 3:1 molar ratio with 20\% Chol).
Thus, the computational phase-field model informed by experiments has a considerable potential to assist in
the design of liposomes with spatially organized surfaces, thereby containing the cost
and time required by the design process.

\section*{Introduction}

In natural membranes, lateral organization of lipids into distinct dynamic entities,
often called lipid rafts or domains, has been recognized as a critical mechanism
for dynamic control of the spatial organization of membrane components.
Lipid rafts are often enriched in sphingolipids (SLs) and cholesterol (Chol),
where the long and saturated acyl chains in SLs enable Chol to intercalate tightly with these lipids,
leading to the formation of liquid ordered phase \cite{brown2000,Lingwood46}.
In contrast, loosely packed phospholipids with unsaturated acyl chains in the membrane form liquid disordered phase.
The difference in packing ability among these lipids results in phase separation and formation of rafts
\cite{Brown1}. Lipid rafts have been linked to a wide range of cellular functions,
from membrane trafficking to inter- and intracellular signaling
\cite{brown2000,Lingwood46,Brown1}.
Membrane phase separation has thus been the focus of intense research in biophysics,
often using model membranes, in the past few decades. One of the most common model
membranes for these studies are giant unilamellar vesicles (GUVs) as their micron-scale
dimensions allows for direct examination of individual vesicles under optical microscopy \cite{Wesolowska2009}.
Therefore, fluorescence-based microscopy techniques have been extensively applied to study raft formation in GUVs. Studies on phase separation in GUVs have shed light on different aspects of membrane domains such as their fluidity
\cite{Kahya2003,SEZGIN20121777}, morphology \cite{B901587F}, coarsening dynamics \cite{Stanich2013},
and thermodynamic equilibria \cite{Veatch17650,FIDORRA20092142}. Moreover, advancement in
image analysis methods have enabled more accurate assessment of lipid raft formation on GUVs,
providing quantitative measurements of the size and shape of these domains
\cite{SEZGIN20121777,FIDORRA20092142,JUHASZ20092541,Bandekar2012}.

With the improved understanding of membrane phase separation,
this phenomenon has further been explored as a tool for the design of liposomes
with heterogeneous and spatially organized surfaces. These liposomes have, for instance,
provided promising platforms for delivery purposes. Pioneered by Sofou and her collaborators
\cite{Kempegowda2009,Bandekar2013,Sempkowski2016},
phase separation was utilized to create liposomes with small surface regions of high concentrations of a specific
lipid along with its attached targeting moiety. These ``patchy'' liposomes showed a significantly higher level of
targeting selectivity compared to their non-patchy counterparts \cite{Bandekar2013,Sempkowski2016}.
Furthermore, a recent study that utilized phase separation to control the distribution of a cationic lipid
on liposomes, demonstrated that this approach can reduce the toxicity of these fusogenic liposomes
to enhance their delivery performance \cite{Imam2017}. Hence, liposomal membranes with well-defined phase
behavior can overcome some of the major challenges in the field of intracellular delivery.
Design and optimization of such liposomes through experimentation can, however, be expensive
and time-consuming and will greatly benefit from computer aided modelling.  Since the targeting
selectively and cellular interactions of these liposomal carriers will depend on spatial and temporal
characteristics of their domains (i.e. size, number, ripening time, etc.), it is important for a
computational approach to deliver not only a qualitative picture, but also accurate \emph{quantitative}
information about the dynamics of the membrane organization. This paper takes a step towards
the design and validation of computational tools that address this need.

It is only in recent years that computational studies of phase separation in lipid bilayers
have emerged as a complement to experimental investigations \cite{Berkowitz}.
Multicomponent vesicles have been studied with different numerical approaches:
molecular dynamics \cite{Marrink_Mark2003,Vries_et_al2004,Berkowitz}, dissipative particle dynamics \cite{Laradji_Sunil2004,VENTUROLI20061,B901866B}, and continuum based models
\cite{Wang2008,Lowengrub2009,sohn2010dynamics,Li_et_al2012,Funkhouser_et_al2014}.
A limit of molecular dynamics simulation is the length and time scales that can be investigated.
Dissipative particle dynamics is a coarse graining technique that allows for a significant
computational speed-up with respect to molecular dynamics, but is still computationally
very expensive depending on the application. Thus, we focus on continuum based models
that offer a more time- and cost-efficient alternative. We consider the diffuse-interface description
of phase separation given by the Cahn--Hilliard (CH) phase--field model \cite{Cahn_Hilliard1958,CAHN1961}.
In \cite{C2CP41274H}, it is shown that with a suitable modification of the Ginzburg--Landau free energy, the
CH model can successfully simulate lipid microdomains. In our work on phase separation
in steady membranes and complex biologically inspired shapes \cite{Yushutin_IJNMBE2019}, we showed that
numerical results from a surface CH equation successfully reproduce the patterns of lipid domains
experimentally observed in \cite{veatch2003separation}. All the above references
dealing with continuum based models have one key limitation: they do not tackle a quantitative validation
of the phase-field model against experimental data. More precisely, the majority of the works
are only concerned with modeling and numerical aspects of lateral phase separation,
 while in \cite{Funkhouser_et_al2014,C2CP41274H} it is shown that numerical results can \emph{qualitatively}
 reproduce certain experimental trends (i.e., trapped coarsening and power law growth of average raft diameter,
 respectively). Here, we present the first quantitative comparison of the surface CH phase-field model
 with experimental data on lateral phase separation in GUVs.

Besides a valid mathematical description, computer modeling requires effective numerical algorithms,
especially if evolution (rather than equilibrium) and uncertainty quantification are of interest.
Although there exists an extensive literature on numerical methods for the CH equation
in planar and volumetric domains (see, e.g., recent publications
\cite{guillen2014second,tierra2015numerical,liu2015stabilized, cai2017error} and references therein),
there were not so many papers where the equations are treated on surfaces exhibiting curvature until recently
\cite{gera2017cahn,jeong2015microphase,greer2006fourth,Funkhouser_et_al2014,du2011finite,garcke2016coupled}.
Benefiting from these new developments we build our simulation tools on a trace finite element method
(FEM) \cite{olshanskii2017trace2} (one of the most flexible numerical approach to handle complex geometries),
adaptive time integration~\cite{gomez2008isogeometric} (necessary to handle a high variation of temporal scales),
and  state-of-the-art iterative solvers for algebraic systems.

Utilizing the above-mentioned numerical methods to solve the CH equation on curved surfaces,
we model lateral phase separation in a multicomponent liposomal membrane. We apply fluorescence
confocal imaging of electrofomed GUVs of a ternary lipid composition at two distinct molar ratios for
validation of this model. We demonstrate that the results of the present simulations on the number
of domains, their ripening dynamics, and their size and shape are in a very good agreement with those from experiments,
but require a thoughtful choice of model parameters. This continuum based model thus has a considerable
potential for the design of liposomes with spatially organized surfaces, thereby greatly reducing the need
for costly and time-consuming experiments.

%\anna{Breakdown of any form of experimental comparison in the computational papers:}
%\begin{itemize}
%\item \anna{\cite{Funkhouser_et_al2014}: numerical results can capture the following experimental trend:
%slowed evolution, or ``trapped coarsening'', in multicomponent lipid vesicles with a high area-to-volume ratio, where minority phase
%domains were found to repel each other instead of coarsening once they had bulged outward from the membrane. No quantification.}
%\item \anna{\cite{gera2017cahn,jeong2015microphase,greer2006fourth,du2011finite,garcke2016coupled}: no lab experiments are mentioned.}
%\item \anna{\cite{C2CP41274H}: In the interfacial
%dynamics of a two-component system two interesting features
%are numerically and experimentally observed: the morphology
%behaves statistically self-similar in time, and the growth rate of
%the characteristic length scale obeys a temporal power law.
%No quantification.}
%\end{itemize}

\section*{Materials and methods}
\vskip -2mm
\subsection*{Experimental approach}
\vskip -2mm
\subsubsection*{Materials}
Lipids 1,2-dioleoyl-sn-glycero-3-phosphocholine (DOPC), 1,2-dipalmitoyl-sn-glycero-3-phosphocholine (DPPC), 1,2-dipalmitoyl-sn-glycero-3-phosphoethanolamine-N- (lissamine rhodamine B sulfonyl) (Rho-PE) were purchased from Avanti Polar Lipids (Alabaster, AL). We purchased the sucrose from VWR (West Chester, PA). Cholesterol was from Sigma Aldrich (Saint Louis, MO) and chloroform from Omnipure (Caldwell, Idaho). All lipid stock solutions were prepared in chloroform. Indium tin oxide (ITO) coated glasses and microscope glass slides were from Thermo Fisher Scientific (Waltham, MA) and coverslips were from Corning Inc. (Corning, NY).
ITO plates were cleaned using chloroform, ethanol and DI water prior to use. Microscope slides and coverslips were cleaned with ethanol and DI water before usage.

\subsubsection*{Preparation of Small Unilamellar Vesicles (SUVs)}
SUVs were prepared using dehydration-rehydration and tip sonication as described in our previous studies
\cite{park2018reconstitution,majd2005hydrogel}.
In brief, a mixture of DOPC, DPPC, Chol at the desired molar ratio (see below) plus 0.6 mol\% Rho-PE %with the desired lipid composition 
was prepared in chloroform. 1 ml of this solution was added into a 5 ml pearl-shaped flask and was dried under vacuum using a rotary evaporator (Hei-Vap, Heidolph, Germany) for $\sim$2 hr. The resultant lipid film was then hydrated using pre-heated DI water (60\textdegree{}C) to a final lipid concentration of 2.5 mg/ml. The milky suspension was then sonicated %above 60\textdegree{}C
using a tip-sonictor (55-Watt Sonicator Q55, Qsonica, Newtown, CT). The procedure of 1 min tip sonication
at 10 Hz with 30 sec resting intervals was applied for about 20 times to produce a clear solution of SUVs. The SUV solution was stored at 4\textdegree{}C and used within 5 days.

Lipid compositions applied here were (i) DOPC:DPPC with a 2:1 molar ratio with 20\% Chol, referred to as 2:1:20\% composition,
and (ii) DOPC:DPPC with a 3:1 molar ratio with 20\% Chol, referred to as 3:1:20\% composition.
Note that both compositions contained Rh-PE to allow for fluorescence microscopy and visualization of lipid domains.

\subsubsection*{Preparation of Giant Unilamellar Vesicles (GUVs)}
We prepared GUVs using the common technique of electroformation originally developed
by Angelova et al. \cite{DC9868100303} with modifications  detailed in our previous studies
\cite{park2018reconstitution,kang2013simple}. Briefly, \SI{25}{\micro\litre} of aqueous dispersion of SUVs was deposited onto each of two
ITO plates as small droplets and left overnight to dry. A thin PDMS frame with tubing was sandwiched between the two
ITO plates to assemble the electroformation chamber. A solution of sucrose at 235 mM was then slowly injected into
the chamber to rehydrate the lipid patches. Next, we placed the device in a 60\textdegree{}C oven to exceed the highest melting
temperature in the lipid composition (in this case DPPC with melting temperature of 41.2\textdegree{}C) where an AC electrical field was applied through copper tapes attached to the ITO plates.
With a frequency of 50 Hz, the electric field was increased to 2 Vpp at rate of 100 mVpp/min and kept for
$\sim$3 hours using a function waveform generator (4055, BK Precision, Yorba Linda, CA). Once formed, GUVs were detached
by decreasing the frequency to 1 Hz for $\sim$30 min.

\subsubsection*{GUV Imaging}
For microscopy, GUVs were collected from the electroformation
chamber through the outlet tubing and $\sim$\SI{10}{\micro\litre}
of the GUV-containing solution was placed on a clean microscope glass slide that
was then covered with a clean coverslip. The edges were sealed with nail polish to immobilize the coverslip.
The sample was reheated on a hot plate
(Hei-Connect, Heidolph, Germany) to 60\textdegree{}C, for at least 5 min before the imaging started.
All the images were acquired using a Zeiss LSM 800 confocal laser scanning microscope (Zeiss, Germany).
%equipped with 4 lasers and multialkali (MA) PMT.
The sample was placed on the microscope stage where
it gradually cooled down to the room temperature, which was monitored and recorded. It should be noted that the time (for image collection) was recorded with time zero considered as when the sample was removed from the hot plate. We applied epi-fluorescence imaging for the initial assessment of GUVs and their lipid domains and applied confocal imaging to further assess the domains on GUVs and quantify their size as described below. Epi-fluorescence
images were collected using a 40X objective with numerical aperture (NA) of 0.95, with 538-562
excitation filter wavelength and 570-640 emission filter wavelength. Confocal images were collected using a 63X oil objective with NA of 1.40 using a 561 nm wavelength laser.
Confocal image slices were collected 
with 0.4-\SI{0.9}{\micro\metre} Z-steps,
depending on the size of the examined GUV, to minimize the time required
for imaging the entire 
vesicle without significant movement of vesicle. Confocal
images were analyzed using ZEN software (ZEN 2.6 lite, Zeiss, 
Germany).
Considering that Rh-PE has shown preferential partitioning into the liquid disordered phase
\cite{KLYMCHENKO201497,PMID:20642452},
we assumed that 
  \begin{wrapfigure}[12]{l}{0.41\textwidth}
\centering
\begin{overpic}[width=.2\textwidth, grid=false]{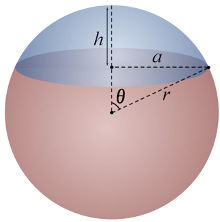}
\put(5,90){\small{A}}
\end{overpic}
\begin{overpic}[width=.2\textwidth, grid=false]{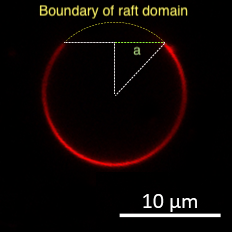}
\put(5,90){\textcolor{white}{\small{B}}}
\end{overpic}
\caption{\small A) Schematic of a vesicle with a spherical cap raft domain (shown in blue). B) Representative confocal image slice of a GUV with a lipid domain (marked with dashed arc).}
\label{fig:cap}
\end{wrapfigure}
dark patches on the examined GUVs represented the liquid ordered phase
while the red regions represented the liquid disordered phase.

\subsubsection*{Image Analysis for Lipid Domain Characterization}
 For the analysis, we assumed that the GUVs were perfect spheres, and
and that the lipid domains were in the form of spherical caps (Fig.~\ref{fig:cap}A). For each GUV,
all collected confocal  image slices were
analyzed to determine the vesicle diameter
(using the confocal slice with the largest circular cross-section of the GUV) and its radius $r$,
to calculate the vesicle total surface area $4 \pi r^2$. For lipid domains,
which corresponded to dark arcs on individual confocal slices (Fig.~\ref{fig:cap}B), diameter
 of the base of the cap was determined from the slice with the largest dark arc for a given domain and
 its corresponding radius $a$ was used to calculate the raft perimeter $2\pi a$.
 Surface area of each cap (i.e.~lipid domain) was calculated using:
\begin{equation}\label{eq:cap}
Area = 2 \pi r^2(1 - \cos \theta),
\end{equation}
where $\theta$ corresponded to the angle shown in Fig.~\ref{fig:cap}A and was calculated using:
%\begin{equation*}
$
\theta = \arcsin \left( \frac{a}{r} \right).
$
%\end{equation*}
%\anna{Sheereen, did Yifei measure the angle $\theta$ or $h$ directly?}

We performed this analysis on 18 GUVs with 2:1:20\% composition and 17 GUVs with 3:1:20\% composition from at least
2 independent experiments per composition.

\subsection*{Computational approach}

\subsubsection*{Mathematical model}
A well established continuum-based model for the process of spinodal decomposition and phase separation
is the CH phase-field model \cite{Cahn_Hilliard1958,CAHN1961}. %Starting from energetic  principles
In order to state the model, let $\Gamma$ be a sphere representing a liposome with a
\SI{10}{\micro\metre} diameter and distributed mass concentrations $c_i = m_i/m$, {\small$i = 1, 2$}.
Here, $m_i$ is the specific masses of each
phase and $m=m_1+m_2$. We choose $c = c_1$, $c\in [0,1]$, to be the representative concentration,
e.g. the concentration of the ordered phase, meaning $c\sim 1$ in ordered phase and $c\sim 0$ in the disordered phase.
The surface CH equation governs the evolution in time $t$ of $c=c(t,\vect x)$, $\vect x \in\Gamma\subset\mathbb R^3$:
\begin{equation}\label{surfaceCH}
\left\{
\begin{split}
\frac{\partial c}{\partial t}& = \nabla_\Gamma\cdot\left(M_c\,\nabla_\Gamma\left(f_0'(c) - \epsilon^2\,\Delta_\Gamma c\right)\right)\quad \text{on}~\Gamma,~\text{for}~t>0,\\
 c &= c_0,\quad \text{at}~t=0,
 \end{split}
\right.
\end{equation}
In \eqref{surfaceCH}, $c_0=c_0(\vect x)$ is an initial distribution of concentration, corresponding to a homogeneous mixture,
$f_0(c) = \frac{1}{4}\,c^2\,(1 - c)^2$
%\anna{(is it like this or is there a coefficient?) We use =1 Since we have $D$ and $\epsilon$, this coefficient is redundant.}
is the specific free energy of a homogeneous phase, %the Ginzburg--Landau double-well potential
$\nabla_\Gamma$ stands for the tangential gradient, and $\Delta_\Gamma$ is the Laplace--Beltrami operator.
See also~\cite{Yushutin_IJNMBE2019}. Problem~\eqref{surfaceCH} is obtained from
minimizing the total specific free energy $\int_\Gamma f_0(c) + \frac{1}{2} \epsilon^2 | \nabla_\Gamma c |^2\,ds$
subject to the conservation of total concentration $\int_\Gamma c\,ds$.
Parameter~$\epsilon > 0$ in the free energy functional defines the width of the (diffuse) interface between the phases.
%, $\lbrack \epsilon \rbrack = \unitLength$. For the mobility function~
Finally, $M_c$ is the so-called mobility coefficient (see \cite{Landau_Lifshitz_1958}). We consider the degenerate mobility of the form
\begin{equation}\label{mobility}
M_c = D c\,(1-c)
\end{equation}
with diffusivity constant~$D > 0$. Mobility \eqref{mobility} is a popular choice for numerical studies.
Although it is known that the dependence between the mobility and the concentration
produces important changes during the coarsening process, only a few authors consider more complex
mobility functions; see, e.g., \cite{PhysRevE.60.3564}. In the absence of studies on the
appropriate mobility function for lateral phase separation in liposomes, here we choose to use \eqref{mobility}.

While both model parameters $D$ and $\epsilon$ correspond to thermodynamics  properties of matter,
their direct evaluation is not straightforward.
In particular, the coefficient $D$ determines the rate of change of %the order parameter
$c$ depending on the specific free energy fluctuations, rather than depending on a molar flux due to molecular Brownian motion. Therefore, the known rates for lateral diffusion in lipid membranes~\cite{lindblom1994nmr} are of limited help in setting $D$.
Given the uncertainties about the values of $D$ and $\epsilon$, we follow an alternative (data driven) approach:  If one looks at \eqref{surfaceCH} as a dynamical system, then the time scale depends linearly on $D$, while  $\epsilon$ defines the relative duration of the (fast) decomposition and (slow) patterns evolution phases.  This allows us to apply backward optimization: we set the coefficient values to match the time evolution of the patterns observed  \textit{in vitro}.
This approach suggested the following values:
$D=10^{-5}(\mbox{cm})^2s^{-1}$ for the 3:1:20\% composition and $D=2.5\,10^{-5}(\mbox{cm})^2s^{-1}$  for the
2:1:20\% composition, and $\epsilon=\SI{0.05}{\micro\metre}$ for both compositions. The value of  $\epsilon$
over-estimates the 5 nm prediction of the transition layer width found in \cite{Risselada2008} because
such width is beyond the current resolution capabilities of the discrete continuum model. Note that the sensible
variation of $D$ with the membrane composition and temperature should be expected
and it (partially) compensates for  the \textit{unknown} dependence of the free energy functional form  on the composition.

In order to model an initially homogenous liposome, the initial concentration $c_0$
is defined  as a realization of Bernoulli random variable~$c_\text{rand} \sim \text{Bernoulli}(a_{\text{raft}})$
with mean value $a_{\text{raft}}$, i.e. we set:
\begin{equation}\label{raftIC}
	c_0 \coloneqq c_\text{rand}(\vect x)\quad\text{for active mesh nodes $\vect x$}.
\end{equation}
Following the thermodynamic principles described in the next section,
we set $a_{\text{raft}}=0.1$ for the 3:1:20\% (DOPC\,:\,DPPC\,:\,Chol) composition and $a_{\text{raft}}=0.16$ for the 2:1:20\%
composition. 

\subsubsection*{Numerical method and input data}\label{sec:num_meth}
With the exception of few equilibrium states, the CH equation lacks analytical solutions.
Thus, one has to resort to a numerical solution.  We discretize problem~\eqref{surfaceCH} with
the trace finite element method (Trace FEM), a state-of-the-art computational technique for
systems of partial differential equations (PDEs) posed on surfaces~\cite{olshanskii2017trace2}.
The first two steps in the application of Trace FEM are common to other finite element methods. First, one
rewrites \eqref{surfaceCH} as the system of two second order PDEs
by introducing the chemical potential as another unknown variable: $\mu = f_0'(c) - \epsilon^2\Delta_\Gamma  c $.
Then, one proceeds to an equivalent integral form of the PDE system, also known in PDE theory (see, e.g.~\cite{evans2010partial})
as weak formulation. The weak formulation is obtained by multiplying the equations by smooth test functions,
integrating the equations over $\Gamma$, and applying the surface Stokes formula. For the
CH equation~\eqref{surfaceCH}, the weak formulations reads: Find
concentration $c$ and chemical potential $\mu$ such that
\begin{align}
&\int_\Gamma \frac{\partial c}{\partial t} \,v \, ds + \int_\Gamma M_c\nabla_\Gamma \mu \,\nabla_\Gamma v \, ds = 0, \label{eq:sys_CH1_weak} \\
&\int_\Gamma  \mu \,q \, ds - \int_\Gamma f_0'(c) \,q \, ds - \int_\Gamma \epsilon^2\nabla_\Gamma c \,\nabla_\Gamma q \, ds = 0, \label{eq:sys_CH2_weak}
\end{align}
for any sufficiently regular test functions $v$ and $q$ on $\Gamma$.

\begin{wrapfigure}{l}{0.24\textwidth}
\centering
\begin{overpic}[width=.24\textwidth, grid=false]{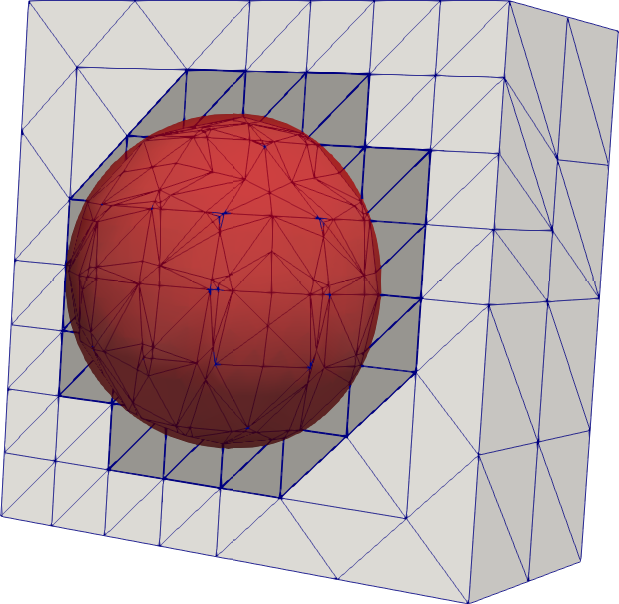}
\end{overpic}
\caption{\small A model liposome (red) immersed in a bulk tetrahedral mesh (gray).}
\label{fig:grid}
\end{wrapfigure}

The remaining steps are specific to Trace FEM.
The sphere $\Gamma$ is immersed in a cube, which is tessellated into tetrahedra.
See Fig.~\ref{fig:grid}.
This tessellation forms a regular triangulation of the bulk domain in the sense of \cite{ciarlet2002finite}. The zero level set of
the $P_1$ (i.e., linear) Lagrangian interpolant (to the vertices of the tetrahedra) of the signed distance function of $\Gamma$ provides a polyhedral approximation $\Gamma_h$ of the sphere, which will further be used for numerical integration instead of $\Gamma$ in \eqref{eq:sys_CH1_weak}--\eqref{eq:sys_CH2_weak}.  Tetrahedra intersected
by $\Gamma_h$ form an active mesh $\mathcal{T}^{\rm bulk}$ that supports the degrees of freedom
(dark gray layer in Fig.~\ref{fig:grid}). On  $\mathcal{T}^{\rm bulk}$ we further define a
(finite dimensional linear) space of continuous functions, which are polynomials of degree~1 on each tetrahedra from $\mathcal{T}^{\rm bulk}$.
Seeking approximations of $c$ and $\mu$ in such space (denoted with $c_h$ and $\mu_h$)
so that that  eq.~\eqref{eq:sys_CH1_weak}--\eqref{eq:sys_CH2_weak}
are satisfied for $v$ and $q$ in the same space reduces problem \eqref{eq:sys_CH1_weak}--\eqref{eq:sys_CH2_weak}
to a large (but finite) system of ordinary differential equations (ODEs). That constitutes the Trace FEM; see \cite{Yushutin_IJNMBE2019,yushutin2020numerical} for further implementation details.
The finite element approximation to the solution enjoys a guaranteed convergence to the true solutions of the
PDE problem if the mesh  $\mathcal{T}^{\rm bulk}$ is refined~\cite{ORG09,elliott1992error}.
Hence, the fidelity of the numerical solution is ensured by a sequence of mesh
refinements until the solutions on two subsequent  meshes demonstrate the same
qualitative and quantitative behavior. The final finest mesh, which is the one we adopted for the numerical results
reported in this paper, was produced by a 6-level refinement of the coarse mesh shown in  Fig.~\ref{fig:grid},
leading to $87,728$ active degrees of freedom.
The resulting system of ODEs is integrated in time for $t=0$ to final time $t^{\rm final}=2.5\,10^{4}s$
using a semi-implicit stabilized Euler method
~\cite{Shen_Yang2010}  and an adaptive time stepping
technique~\cite{gomez2008isogeometric}. Thus, the piecewise polynomial approximations $c_h$ and $\mu_h$
become available at times $t_n\in[0, t^{\rm final}]$, $n=1,\dots,N$. The time step $\Delta t=t_n-t_{n-1}$ adaptively varies
from  $\Delta t=2.5 \cdot10^{-5}$ s during the fast initial phase of spinodal decomposition to about $\Delta t=0.1-10$ s
during the later slow phase of rafts coarsening and growth, and up to $\Delta t=10^3$ s when the process is close to equilibrium.
To pass from time $t_{n-1}$ to time $t_{n}$, a system of algebraic equations needs to be solved in order
to get $c_h$ and $\mu_h$ at time $t_{n}$.
Such system features a large sparse matrix that has a block structure. To solve it,
the GMRES~\cite{saad1986gmres} iterative method with a block preconditioner (see, e.g.,
\cite{benzi_golub_liesen_2005}) is successfully applied.

Note that the finite element method conforms to the mass conservation principle behind \eqref{eq:sys_CH1_weak}--\eqref{eq:sys_CH2_weak}
and hence the numerical solution satisfies
\begin{equation}\label{raftFracDiscrete}
\int_{\Gamma_h} c_h(\vect x, t_n) \diff{s}=\int_{\Gamma_h} c_h(\vect x, t_{n-1}) \diff{s}\quad\text{implying}\quad	\frac{\int_{\Gamma_h} c_h(\vect x, t_n) \diff{s}}{\int_{\Gamma_h} 1 \diff{s}} \simeq a_{\text{raft}}
\end{equation}
for all $n=1,\dots,N$. Another quantity of interest, is the total perimeter of the rafts, $p_{\text{raft}}$, which can be defined as the length of the (multi-component) curve that is the level set $c=1/2$. This definition is implicit, so for numerical purposes we set
\begin{equation}\label{perimeter}
p_{\text{raft}}(t_n):=p_0\int_{\Gamma_h}\epsilon|\nabla_\Gamma c_h(\vect x, t_n)|^2 \diff{s}.
\end{equation}
For small $\epsilon$, this quantity converges to the length of the level-set $c=1/2$.
Here $p_0$ is a calibration constant such that the computed perimeter  is exactly $2\pi$ for the equilibrium solution
of the 1:1 composition.

\section*{Results and Discussion}

\begin{figure}[h]
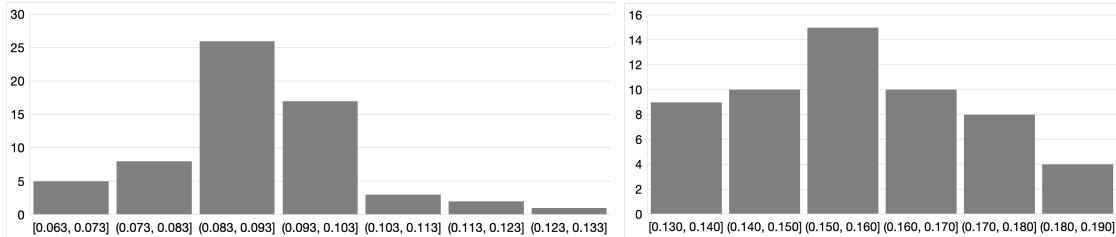

\centering
	\begin{subfigure}{.49\textwidth}
		\centering
		\includegraphicsw{{raft_area_fraction_311}.png}
	\end{subfigure}
		\begin{subfigure}{.4\textwidth}
		\centering
		\includegraphicsw{{raft_area_fraction_2120}.png}
	\end{subfigure}%
%\quad
%\begin{tabular}{|c|c|}
%			\hline
%			average  & 0.0904 \\
%			\hline
%			standard deviation & 0.0116\\
%			\hline
%			\end{tabular}
	\caption{Left: Distribution of experimental measurements of the raft area fraction, with average 0.09 and standard deviation 0.011,
	 for composition 3:1:20\%. The total number of measurements is 62 and they are related to 17 GUVs.
	 Right: Distribution of experimental measurements of the raft area fraction with average 0.157 and standard deviation 0.015, for
	composition 2:1:20\%. The total number of measurements is 56 and they are related to 16 GUVs.}
	\label{fig:raft_area}		
\end{figure}

\begin{figure}[h]
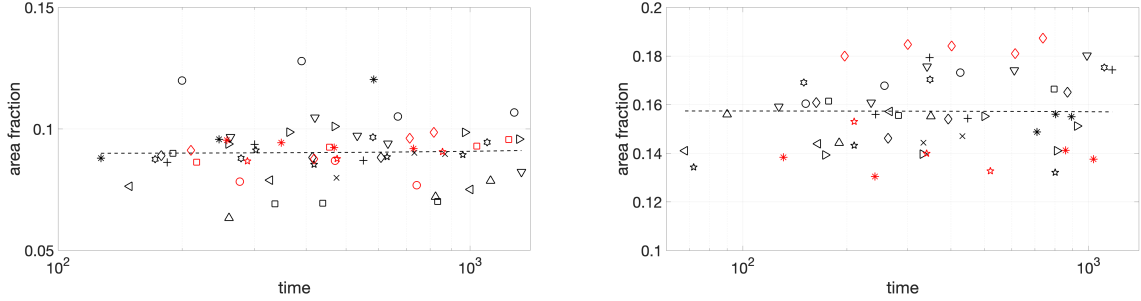

	\begin{subfigure}{.49\textwidth}
		\centering
		\includegraphicsw{{area_fraction_10}.png}
	\end{subfigure}%
\begin{subfigure}{.49\textwidth}
		\centering
		\includegraphicsw{{area_fraction_16}.png}
	\end{subfigure}%	
	\caption{Experimentally measured raft area fraction over time for composition 3:1:20\% (left) and 2:1:20\% (right). Different markers correspond to different GUVs.}
	\label{figS}		
\end{figure}

Ternary membranes composed of a lipid with low transition temperature ($T_m$),
a lipid with high $T_m$, and a sterol such as cholesterol, have frequently been reported to separate
into two co-existing phases of liquid ordered ($l_o$) and liquid disordered ($l_d$) near room temperature
when mixed in proper ratios \cite{PMID:20642452,veatch2003separation}.
One such membrane composition is DOPC:DPPC:Chol, in which $l_d$ phase is composed primarily of DOPC
(lipid with unsaturated acyl chains and low $T_m$) and $l_o$ phase is primarily composed of Chol and DPPC
(lipid with saturated acyl chains and high $T_m$). Here, we prepared two sets of GUVs with this ternary membrane
composition at molar ratios of 2:1:20\% and 3:1:20\% using electroformation.
We studied phase separation on these GUVs at 19.2\textdegree{}C for 2:1:20\%  and 17.5\textdegree{}C for 3:1:20\% composition,
as we expected the latter to have a lower miscibility temperature \cite{veatch2003separation}.
Trace amounts of Rh-PE in GUVs enabled monitoring phase separation using fluorescence microscopy,
where this lipid partitioned into $l_d$ phase, providing a great contrast between the two phases.
Using confocal fluorescence microscopy, we examined a minimum of 17 GUVs (from 2-3 independent experiments)
for the number of their lipid domains as well as area and perimeter of domains at different time points, for each GUV composition.
Fraction of vesicle surface area occupied by dark domains, i.e.~raft area fraction,
in GUVs was then calculated and is summarized in Fig.~\ref{fig:raft_area}.
Histograms in Fig.~\ref{fig:raft_area} show the distribution of raft area fractions for compositions 3:1:20\% and 2:1:20\%.
%, respectively, along with their corresponding average and standard deviation.
Notably, this area fraction showed only slight (i.e. non-significant) changes at different time points
on a given GUV while the number of domains reduced with time (see Fig.~\ref{figS}).
To validate our experimental results, we compared these results to area fractions predicted
by the literature-reported phase diagrams for this ternary membrane.

%\begin{figure}[h]
%	\centering
%	\begin{subfigure}{.55\textwidth}
%		\centering
%		\includegraphicsw{{raft_area_fraction_2120}.png}
%	\end{subfigure}%
%\quad
%\begin{tabular}{|c|c|}
%			\hline
%			average  & 0.1571 \\
%			\hline
%			standard deviation & 0.0152\\
%			\hline
%			\end{tabular}
%	\caption{Distribution of experimental measurements of the raft area fraction and corresponding average and standard deviation for
%	composition 2:1:20\%. The total number of measurements is 56 and they are related to 16 GUVs.}
%	\label{fig:raft_area:16}		
%\end{figure}

From a thermodynamic point of view, co-existence of different phases in ternary
membranes of DOPC:DPPC:Chol at equilibrium can be described through phase diagrams.
These diagrams suggest that at temperatures studied here (16-19\textdegree{}C),
DOPC:DPPC:Chol membrane, at both molar ratios of 3:1:20\% and 2:1:20\%,
is in a binary state of $l_d$ and $l_o$ coexistence \cite{Veatch17650}.
Given the sensitivity of phase diagrams to temperature, we focused on those closest to our experimental temperatures,
i.e.~20\textdegree{}C for 2:1:20\% and 15\textdegree{}C for 3:1:20\% composition.
Relying on tie-lines and their corresponding endpoints on the phase diagrams \cite{Veatch17650},
we determined the lipid composition of the two liquid
phases in the above-mentioned membranes (see Table 1). Using these values, we then
proceeded to find the fraction of total lipids in each phase and finally, the raft area fraction for
each membrane composition, as described below and summarized in
Table \ref{tab:thermo}.

\begin{table}[ht]
\centering
\begin{tabular}{ |c|c|c|c|c|c|c|c|c|c|}
 \hline
 \footnotesize{Membrane composition} & \multicolumn{3}{| c |}{\footnotesize{Liquid ordered ($l_o$)}}  & \multicolumn{3}{| c |}{
 \footnotesize{Liquid disordered ($l_d$)}} &  \multicolumn{2}{| c |}{\footnotesize{Lipid fraction}} & \footnotesize{Area fraction} \\
 \hline
 \footnotesize{DOPC:DPPC:Chol (Temp)} & \footnotesize{DOPC} & \footnotesize{DPPC} & \footnotesize{Chol} & \footnotesize{DOPC} & \footnotesize{DPPC} & \footnotesize{Chol} & \footnotesize{$\alpha_{l_o}$} & \footnotesize{$\alpha_{l_d}$} & \footnotesize{$a_{\text{raft}}$}\\
  \hline
 \footnotesize{3:1:20\% (15\textdegree{}C)} & \footnotesize{22\%} & \footnotesize{43\%} & \footnotesize{35\%} & \footnotesize{66\%} & \footnotesize{16\%} & \footnotesize{18\%} & \footnotesize{0.13} & \footnotesize{0.86} & \footnotesize{10.8}\% \\
 \hline
 \footnotesize{2:1:20\% (20\textdegree{}C)} & \footnotesize{20\%} & \footnotesize{48\%} & \footnotesize{32\%} & \footnotesize{60\%} & \footnotesize{22\%} & \footnotesize{18\%} & \footnotesize{0.17} & \footnotesize{0.83} & \footnotesize{13.6\%} \\
  \hline
\end{tabular}
\caption{Phase composition and fractions according to phase diagram analysis.}\label{tab:thermo}
\end{table}%

The fraction of lipids in each of the coexisting phases can then be calculated by \cite{DAVIS2009521}:
\begin{equation*}
\alpha_{l_i} = \frac{X^{\text{Lipid}}_{l_j} - X^{\text{Lipid}}}{X^{\text{Lipid}}_{l_j} - X^{\text{Lipid}}_{l_i}},
\end{equation*}
where $\alpha_{l_i}$ ($i = d, o$) is the fraction of lipids that are in the corresponding phase $l_i$, $X^{\text{Lipid}}$ represents
the molar fraction of a specific lipid in the membrane, $X^{\text{Lipid}}_{l_i}$ represents the molar fraction of this lipid
in the corresponding phase, and $X^{\text{Lipid}}_{l_j}$ ($j= d, o$ and $j \neq i$) is the molar fraction of this lipid in the other phase.
Given the two-phase state of the membrane, the remaining of lipids are in the other phase, and thus:
\begin{equation*}
\alpha_{l_j} = 1 - \alpha_{l_i}.
\end{equation*}
See Table \ref{tab:thermo} for the summary of lipid fractions.

The area occupied by each of the phases can be approximated based on the number
of lipid molecules and their corresponding cross-sectional areas in each phase.
Cholesterol is known to intercalate with the tail region of its surrounding lipids,
exerting a condensation effect on these lipids \cite{PhysRevE.80.021931,kheyfets2015area,Edholm2005}.
%\anna{Sheereen, I only found one of the paper in arXiv. Has it appeared anywhere?} {\color{red}I think it's OK to cite arXiv.}
%\anna{Sure, my only concern is that it's a 2015 preprint and it looks kind of weird if it hasn't been accepted in a journal.}
This effect has been studied for DOPC bilayers and DPPC bilayers independently.
In the case of DOPC/Chol mixtures, the area per lipid molecule (DOPC and Chol)
was shown to decrease, with an approximately linear relationship, with increasing Chol content
(up to 50\%) \cite{PhysRevE.80.021931}.
Assuming a molecular area of \SI{72}{\angstrom}$^2$ for DOPC lipid (in the absence of Chol) at 30\textdegree{}C,
and an area contraction coefficient of \SI{0.14}{\angstrom}$^2$/\textdegree{}C for this lipid \cite{KUCERKA20112761}
the lipid molecular area for DOPC/Chol in a particular phase $\left(A_{l_i}^{\text{DOPC/Chol}}\right)$ at temperature $T$ can be calculated by:
\begin{equation*}
A_{l_i}^{\text{DOPC/Chol}} = 72 - (47.5 X^{\text{Chol}}_{l_i}) - 0.14(30 - T),
\end{equation*}
where $X^{\text{Chol}}_{l_i}$ is Chol mole fraction within the corresponding phase. For DPPC,
increasing Chol content has been reported to reduce the area per lipid molecule (DPPC and Chol)
in a non-linear fashion \cite{Edholm2005,kheyfets2015area}.
Assuming an area contraction coefficient of \SI{0.19}{\angstrom}$^2$/\textdegree{}C for DPPC \cite{KUCERKA20112761},
the area per lipid molecule for DPPC/Chol in a particular phase $\left(A_{l_i}^{\text{DPPC/Chol}}\right)$ at temperature $T$
can be found by:
\begin{equation*}
A_{l_i}^{\text{DPPC/Chol}} = A^{\text{DC}}(X^{\text{Chol}}_{l_i}) - 0.19(25 - T),
\end{equation*}
where $A^{\text{DC}}(X^{\text{Chol}}_{l_i})$ is the lipid molecular area at the Chol molar fraction of $X^{\text{Chol}}_{l_i}$ at 25\textdegree{}C
from the analytical calculation provided by \cite{kheyfets2015area}.

The number of lipid molecules in a given phase for different lipid species $\left(N_{l_i}^{\text{Lipid}}\right)$ can be calculated by \cite{Khadka2015}:
\begin{equation*}
N_{l_i}^{\text{Lipid}} = N \alpha_{l_i} X^{\text{Lipid}}_{l_i} ,
\end{equation*}
where $N$ represents the total number of lipids, $X^{\text{Lipid}}_{l_i}$ is the molar fraction of this lipid in the corresponding phase.
Assuming Chol is distributed uniformly within each phase, the area occupied by each phase $\left(A_{l_i}\right)$
can be estimated by:
\begin{equation*}
A_{l_i} = \frac{A_{l_i}^{\text{DOPC/Chol}} N_{l_i}^{\text{DOPC}} + A_{l_i}^{\text{DPPC/Chol}} N_{l_i}^{\text{DPPC}}}{1-X^{\text{Chol}}_{l_i}}
\end{equation*}
where $N_{l_i}^{\text{DOPC}}$ and $N_{l_i}^{\text{DPPC}}$ represent the number of DOPC lipids and DPPC lipids, respectively,
in the corresponding phase. Therefore, the raft area fraction (i.e.~the area fraction of $l_o$)
on liposomes of these compositions is:
\begin{equation*}
a_{\text{raft}} = \frac{A_{l_o}}{A_{l_o}+A_{l_d}}.
\end{equation*}

See Table \ref{tab:thermo} for calculated area fractions. Note that raft area fractions are sensitive to temperature,
due to the temperature sensitivity of the membrane phase behavior, and they increase with a decrease in temperature.
For instance, using the same approach to estimate the raft area fraction in 2:1:20\% membrane at 17.5\textdegree{}C
led to a value of 19.9\% for area fraction compared to 13.6\% at 20\textdegree{}C.
The linear interpolation to predict the raft area fraction at 19.2\textdegree{}C between the two above-mentioned fractions
at 17.5\textdegree{}C and 20\textdegree{}C gives the fraction value of 15.6\%.
This value is very close to the experimental average fraction of 15.7\% measured at 19.2\textdegree{}C (see caption
of Fig.~\ref{fig:raft_area}). Similarly, for 3:1:20\% composition, the above estimation
provides a fraction value of 14.9\% at 10\textdegree{}C compared to 10.8\% at 15\textdegree{}C.
The linear extrapolation of these two fractions, predicts a value of 8.7\% for raft area fraction at 17.5\textdegree{}C,
which is again in agreement with our experimental average fraction of 9.0\% (see caption
of Fig.~\ref{fig:raft_area}).
 We thus used area fractions of 16\% for 2:1:20\% composition and 10\% for 3:1:20\% composition
 to set the initial membrane composition in our simulations.

Independent of the experimental results,10 numerical simulations were run for each composition.
All the simulated liposomes had a \SI{10}{\micro\metre} diameter
and they differed in the initial composition, set using the thermodynamics based estimation
outlined earlier.
The total raft perimeter and the total number of rafts were tracked over time for each simulation
and the results were compared to those from experiments. 
The simulated raft area fraction remains constant over time, and hence is not reported alone,
since the CH model is conservative, as we already remarked. It is noteworthy that our experimental results on raft area fraction supported this assumption (see Fig.~\ref{figS}).

\begin{figure}[h]
	\begin{subfigure}{.49\textwidth}
		\centering
		\includegraphicsw{{Fit_10_bis}.png}
	\end{subfigure}%
%	\centering
%\begin{tabular}{|c|c|}
%			\hline
%			power curve coefficients  & value \\
%			\hline
%			$c$ & 53 \\
%			\hline
%			$t_0$ & 21.1 \\
%			\hline
%			$b$ & -0.37 \\
%			\hline
%			\end{tabular}
	\begin{subfigure}{.49\textwidth}
		\centering
		\includegraphicsw{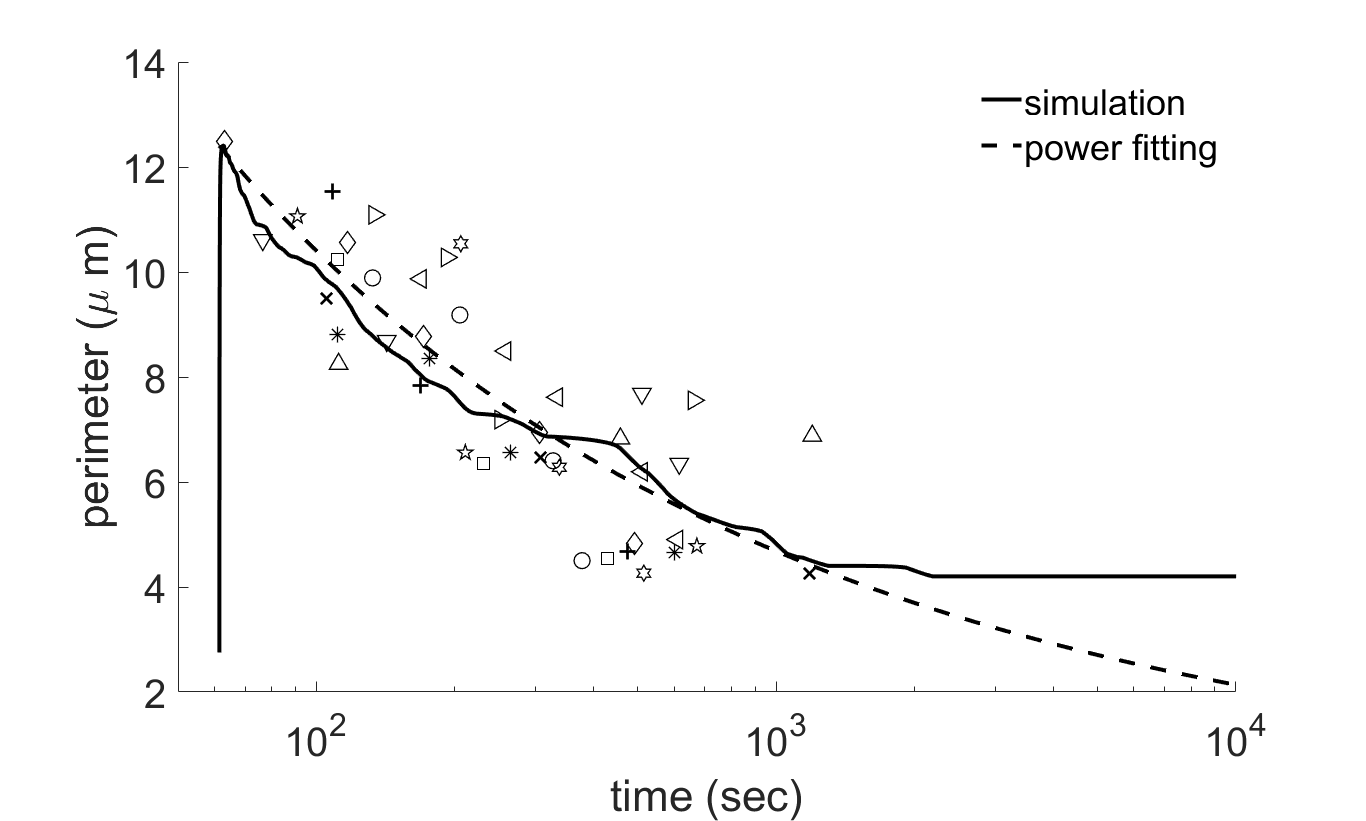}
	\end{subfigure}%
%		\centering
%\begin{tabular}{|c|c|}
%			\hline
%			power curve coefficients  & value \\
%			\hline
%			$c$ & 49 \\
%			\hline
%			$t_0$ & 4.69 \\
%			\hline
%			$b$ & -0.34 \\
%			\hline
%			\end{tabular}
	\caption{Left: Total raft perimeter over time for composition 3:1:20\%: numerical results average (solid line),
	 power curve fitting (dashed line), experimental data (markers). The exponent $b$ in the power curve fitting  \eqref{eq:power_fit}
	 is -0.37.
	 Right: Total raft perimeter over time for composition 2:1:20\%: numerical results average (solid line),
	 power curve fitting (dashed line), experimental data (markers). The exponent $b$ in the power curve fitting  \eqref{eq:power_fit}
	 is -0.34.}
	\label{fig:perimeter}		
\end{figure}

In order to compare the total raft perimeter between simulations and experiments,
we first scaled all the data by the radius of the corresponding GUV, since the diameter of GUVs varied in the experiments
(between 8-\SI{17}{\micro\metre}) while it was constant in the simulations.
Fig.~\ref{fig:perimeter} reports all the rescaled experimental
measurements with markers (a different marker for each GUV) and the average of the computed total raft perimeter
from all the simulations with a solid line for compositions 3:1:20\% and 2:1:20\%. In both cases,
the average of the computed total raft perimeters falls within the cloud of experimental measurements.
For further insights, we then fitted the experimental data with a power function:
\begin{equation}\label{eq:power_fit}
c (t-t_0)^b,
\end{equation}
where exponent $b$ is the critical parameter. The least squares fitting gives $b = -0.37$
for composition 3:1:20\% and $b = -0.34$ for composition 2:1:20\%. The corresponding power
curves are reported in dashed line in Fig.~\ref{fig:perimeter}.

It is known that the limiting behavior of the CH problem satisfies the Mullins--Sekerka dynamics,
characterized by an increase of the average size of (small) rafts and a reduction in the number of rafts.
This process takes the name of Ostwald ripening~\cite{ostwald1900vermeintliche}. For the Ostwald ripening in two
dimensions, it was shown (e.g., in \cite{alikakos2004ostwald}) that a properly defined average of the raft diameters
obeys the well-known $t^{\frac13}$ growth law
predicted in higher dimensions by the Lifschitz--Slyozov--Wagner theory~\cite{bray2002theory}. Hence,
if the dynamics of raft growth in lipid membrane can be quantitatively described by a Ginzburg-Landau type of model,
then the averaged total raft perimeter in the middle phase of coarsening should decay nearly according to $t^{-\frac13}$ law.
This is very close to what we obtained with the least square fitting shown in Fig.~\ref{fig:perimeter},
supporting the common expectation about applicability of CH and similar models for describing phase separation
in lipid membranes. The least squares fitting of the experimental data reinforces our model choice and confirms that the
numerical simulations in average correctly reproduce the trend of the experimental data for both membrane copositions.
Indeed, the solid line (simulation) and the dashed line (power curve fitting) in Fig.~\ref{fig:perimeter}
are close to each other in the the middle phase of raft coarsening, i.e.~roughly between $10^2$ and $10^3$ s.
In particular, we remark a great match for composition 2:1:20\%, whose data give an exponent $b$ in \eqref{eq:power_fit}
closer to $1/3$.

%\begin{figure}[h]
%	\begin{subfigure}{.65\textwidth}
%		\centering
%		\includegraphicsw{Fit_16_perim_rescaled.png}
%	\end{subfigure}%
%		\centering
%%\begin{tabular}{|c|c|}
%%			\hline
%%			power curve coefficients  & value \\
%%			\hline
%%			$c$ & 49 \\
%%			\hline
%%			$t_0$ & 4.69 \\
%%			\hline
%%			$b$ & -0.34 \\
%%			\hline
%%			\end{tabular}
%	\caption{Total raft perimeter over time for composition 2:1:20\%: numerical results average (solid line),
%	 power curve fitting (dashed line), experimental data (markers). The exponent $b$ in the power curve fitting  \eqref{eq:power_fit}
%	 is -0.34. %Right: coefficients for the power curve fitting  \eqref{eq:power_fit}.
%	 }
%	\label{fig:perimeter:16}		
%\end{figure}

We should point out that collecting experimental data from confocal images proved challenging at
early time points due to the rapid changes in domains on GUV surfaces.
In Fig.~\ref{fig:perimeter} we see that the solid line reaches a plateau after $10^3$ s,
indicating that all (or the vast majority) of the simulated liposomes have reached a stable equilibrium. %\anna{Have all the simulations reached the 1-raft stage?}

\begin{figure}[h]
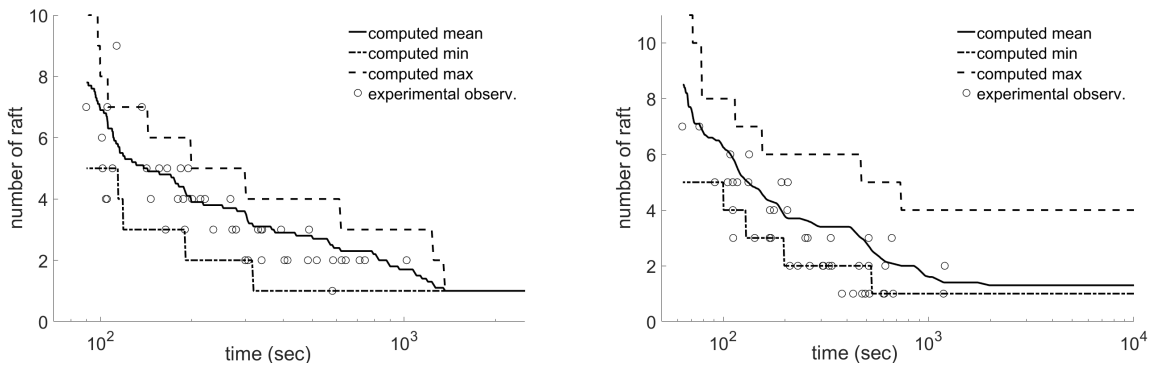

	\begin{subfigure}{.49\textwidth}
		\centering
		\includegraphicsw{{RaftNum10}.png}
	\end{subfigure}%
\begin{subfigure}{.49\textwidth}
		\centering
		\includegraphicsw{{Fit_16_raft_rescaled}.png}
	\end{subfigure}%	
	\caption{Total number of rafts over time for composition 3:1:20\% (left) and 2:1:20\% (right): numerical results average (solid line),
	 minimal and maximum values found numerically (dash-dotted and dashed lines, respectively), and experimental data (circles). }
	\label{fig:Raft}		
\end{figure}

The last quantity we consider is the number of rafts. Fig.~\ref{fig:Raft} reports the experimentally
measured and numerically computed total number of rafts over time for both compositions.
In particular, Fig.~\ref{fig:Raft} shows all the measurements and the numerical results average,
together with the minimum and maximum number of rafts found in the simulations. We observe that the vast majority
of the experimental data (62 measurements for composition 3:1:20\% and 56 for composition 2:1:20\%) falls within
the computed extrema. This is particularly true for composition 3:1:20\%: only three measurements are outliers.
This is further evidence that our simulations based on the CH model
capture the evolution of rafts in lipid membranes well.

\begin{figure}
\begin{center}
\href{https://youtu.be/553F4Z8aauw}{
\begin{overpic}[width=.2\textwidth,grid=false]{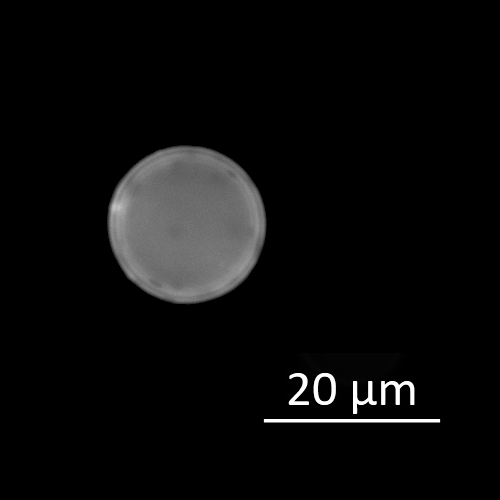}
\put(37,102){\small{$t = 156$}}
\end{overpic}
\begin{overpic}[width=.2\textwidth,grid=false]{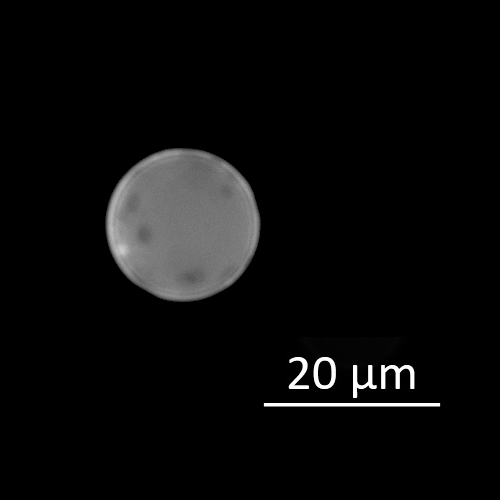}
\put(37,102){\small{$t = 194$}}
\end{overpic}
\begin{overpic}[width=.2\textwidth,grid=false]{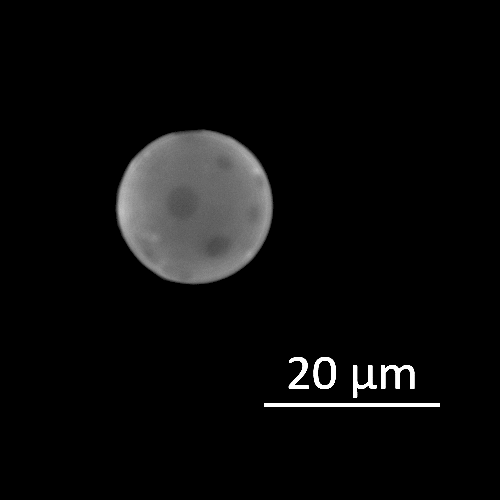}
\put(37,102){\small{$t = 241$}}
\end{overpic}
\begin{overpic}[width=.2\textwidth,grid=false]{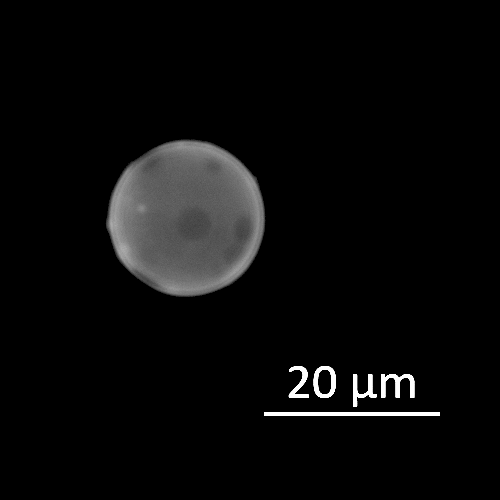}
\put(37,102){\small{$t = 269$}}
\end{overpic}
\\
\begin{overpic}[width=.2\textwidth,grid=false]{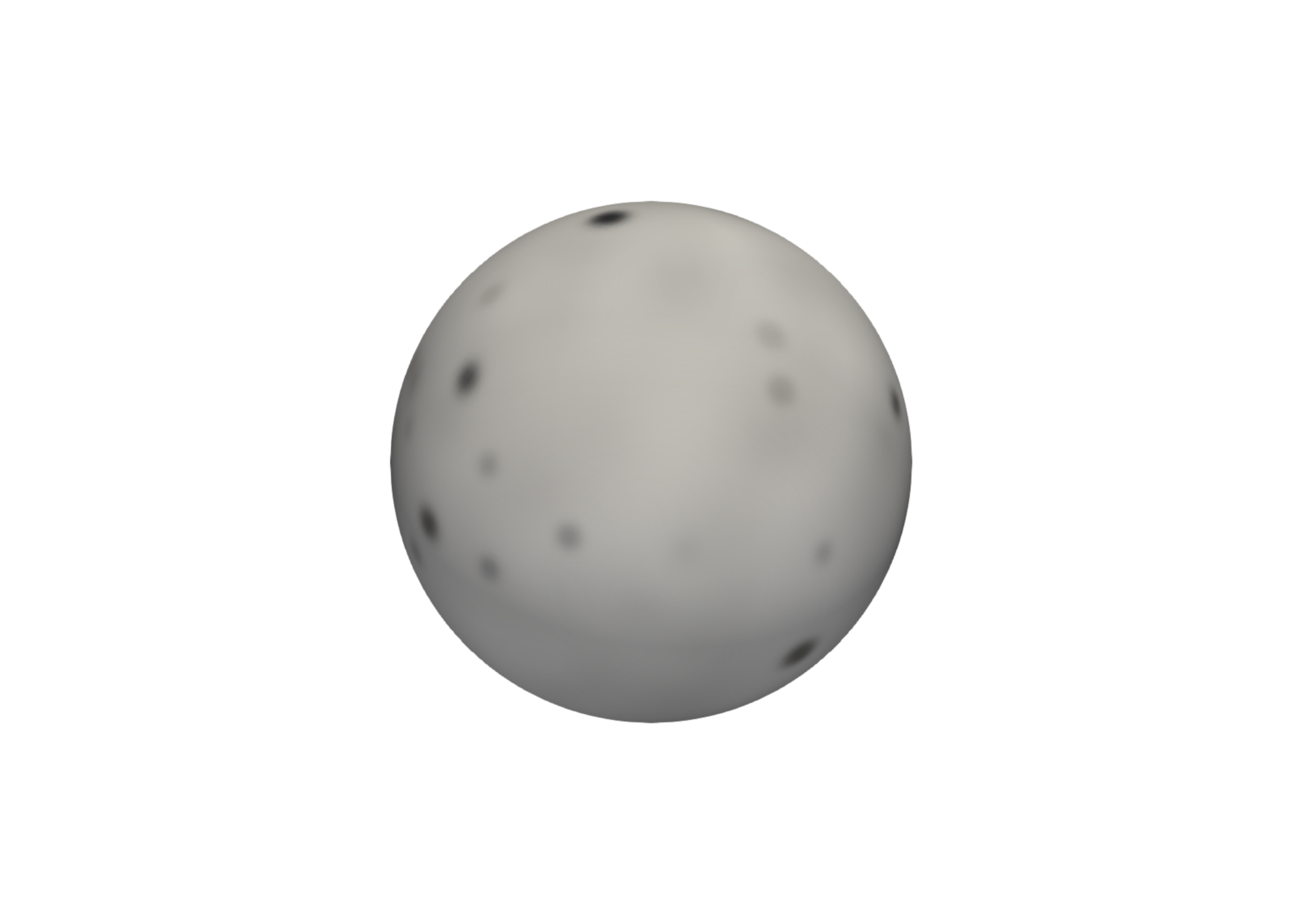}
\end{overpic}
\begin{overpic}[width=.2\textwidth,grid=false]{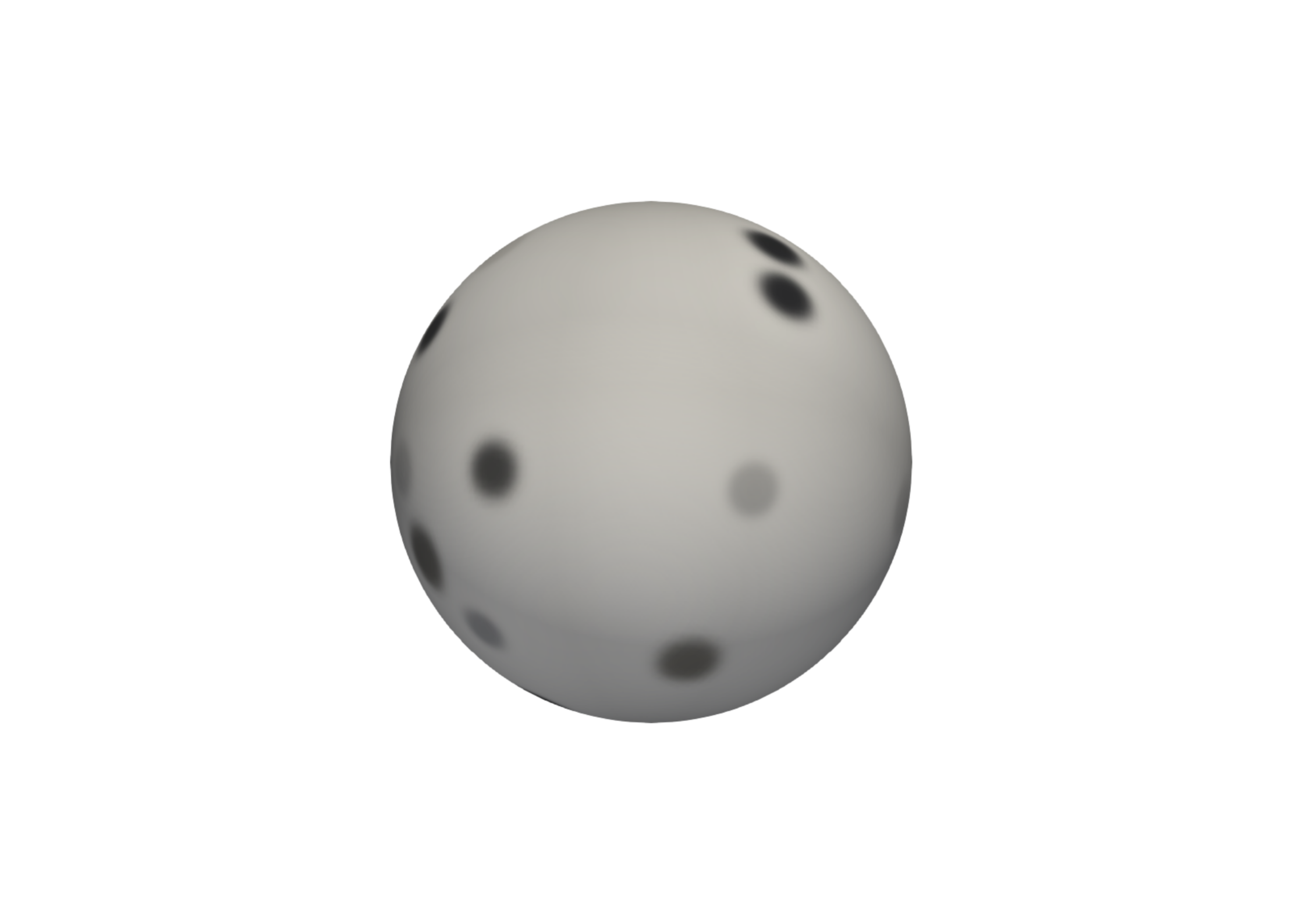}
\end{overpic}
\begin{overpic}[width=.2\textwidth,grid=false]{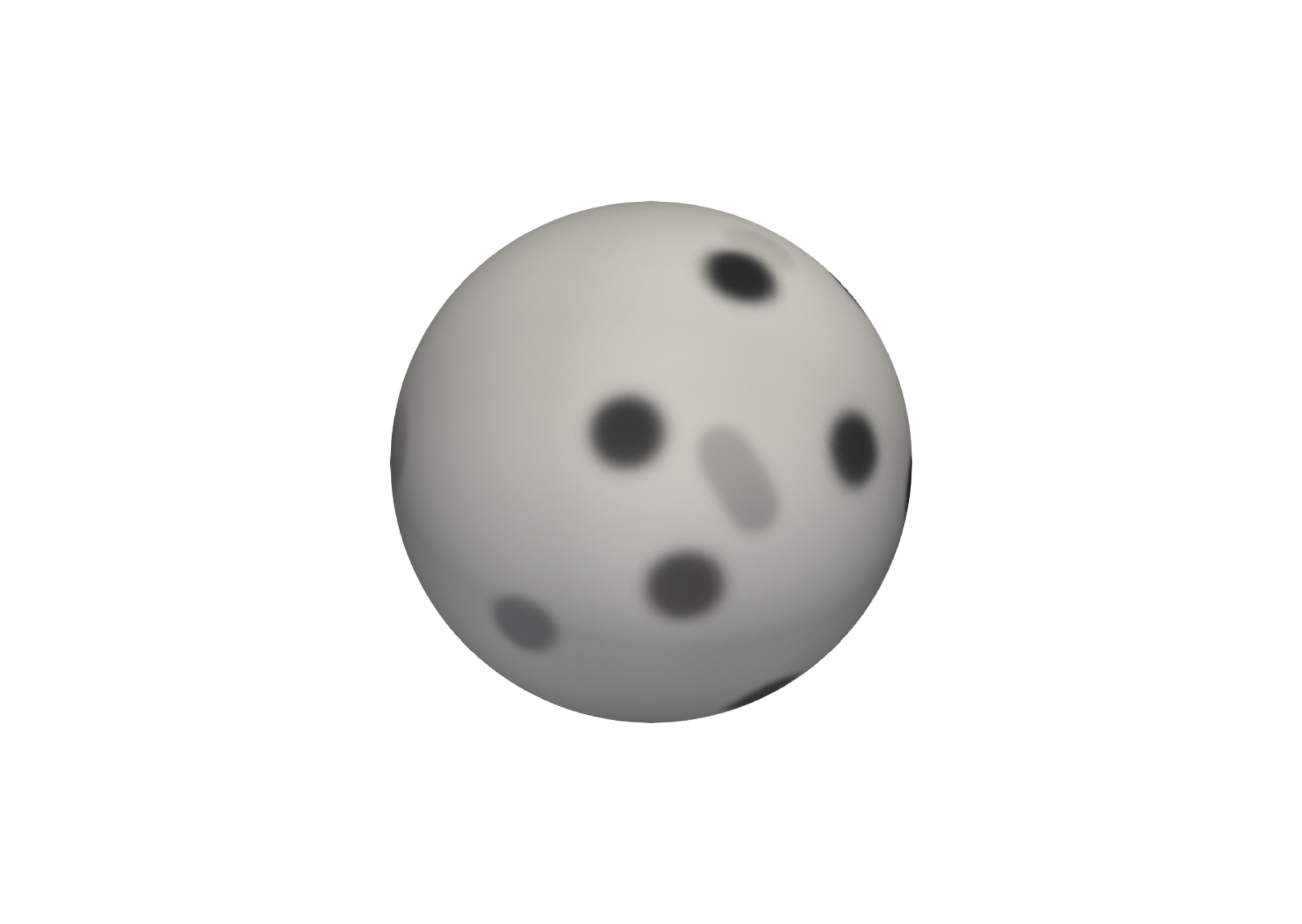}
\end{overpic}
\begin{overpic}[width=.2\textwidth,grid=false]{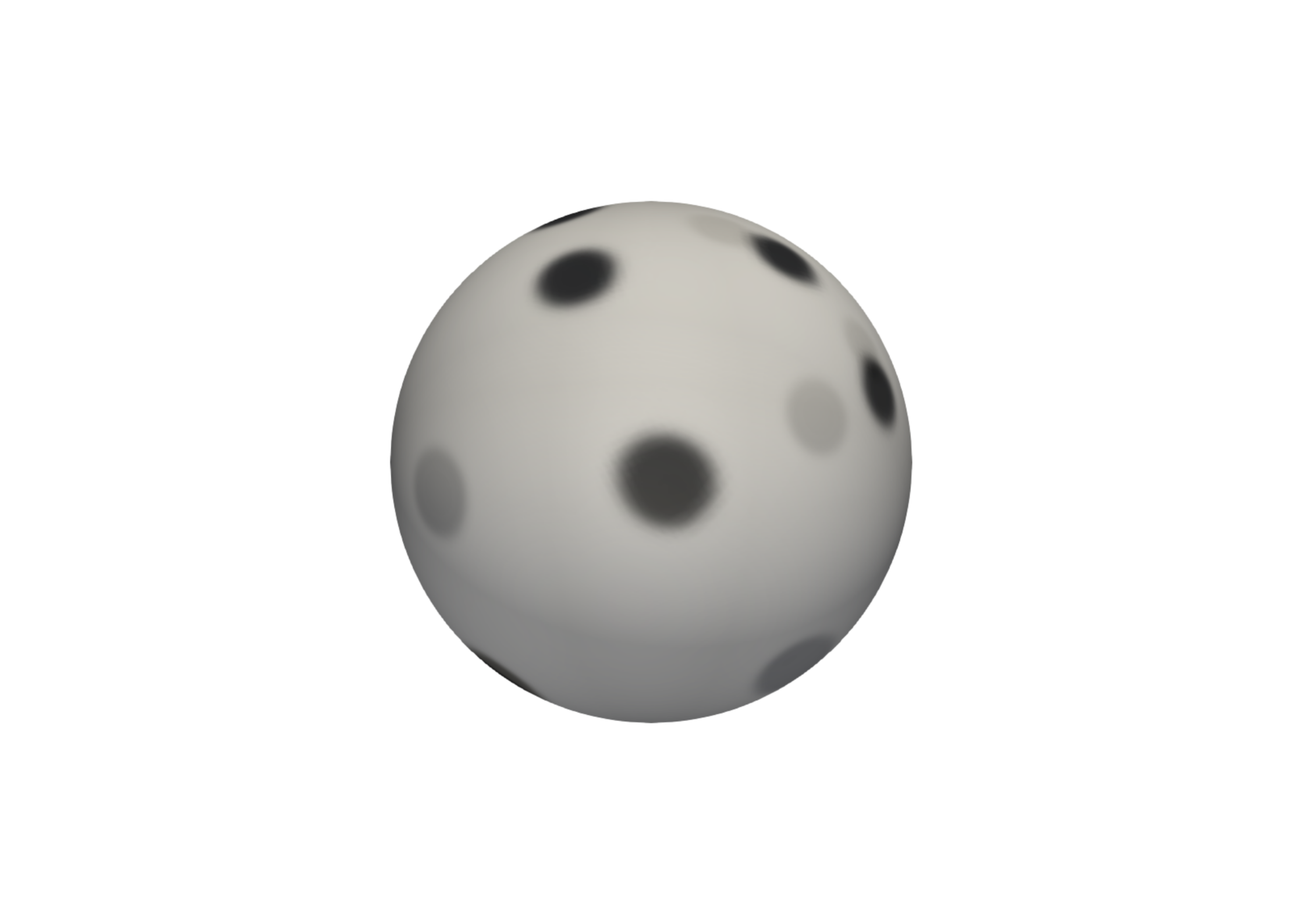}
\end{overpic}
\\
\vskip .5cm
\begin{overpic}[width=.2\textwidth,grid=false]{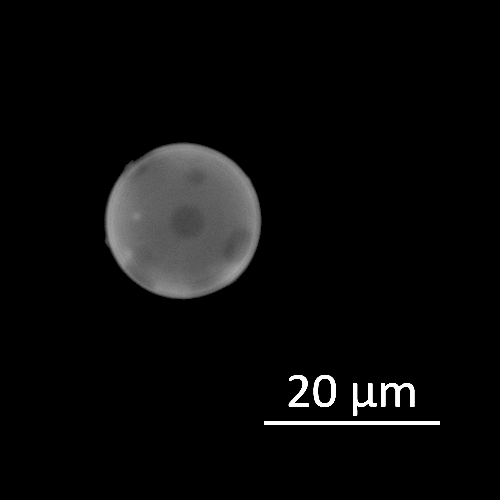}
\put(37,102){\small{$t = 274$}}
\end{overpic}
\begin{overpic}[width=.2\textwidth,grid=false]{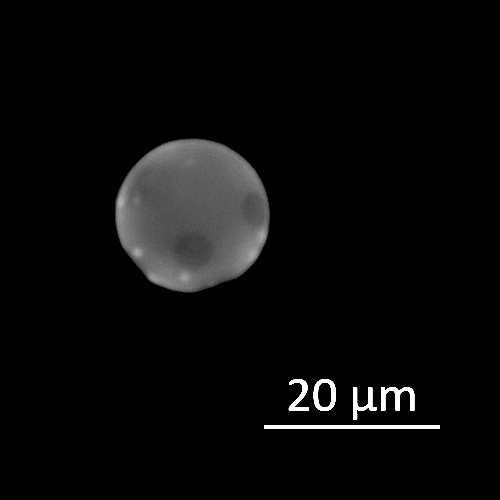}
\put(37,102){\small{$t = 370$}}
\end{overpic}
\begin{overpic}[width=.2\textwidth,grid=false]{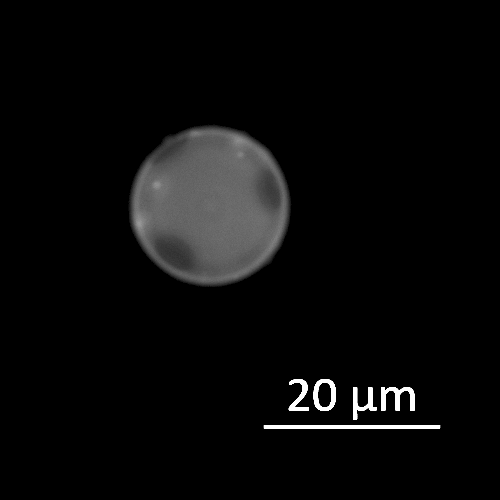}
\put(37,102){\small{$t = 583$}}
\end{overpic}
\begin{overpic}[width=.2\textwidth,grid=false]{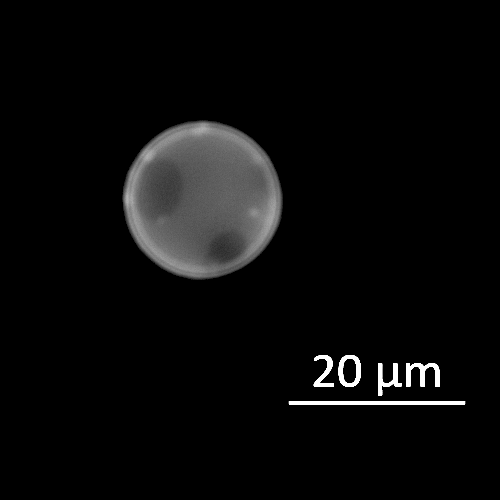}
\put(37,102){\small{$t = 800$}}
\end{overpic}
\\
\begin{overpic}[width=.2\textwidth,grid=false]{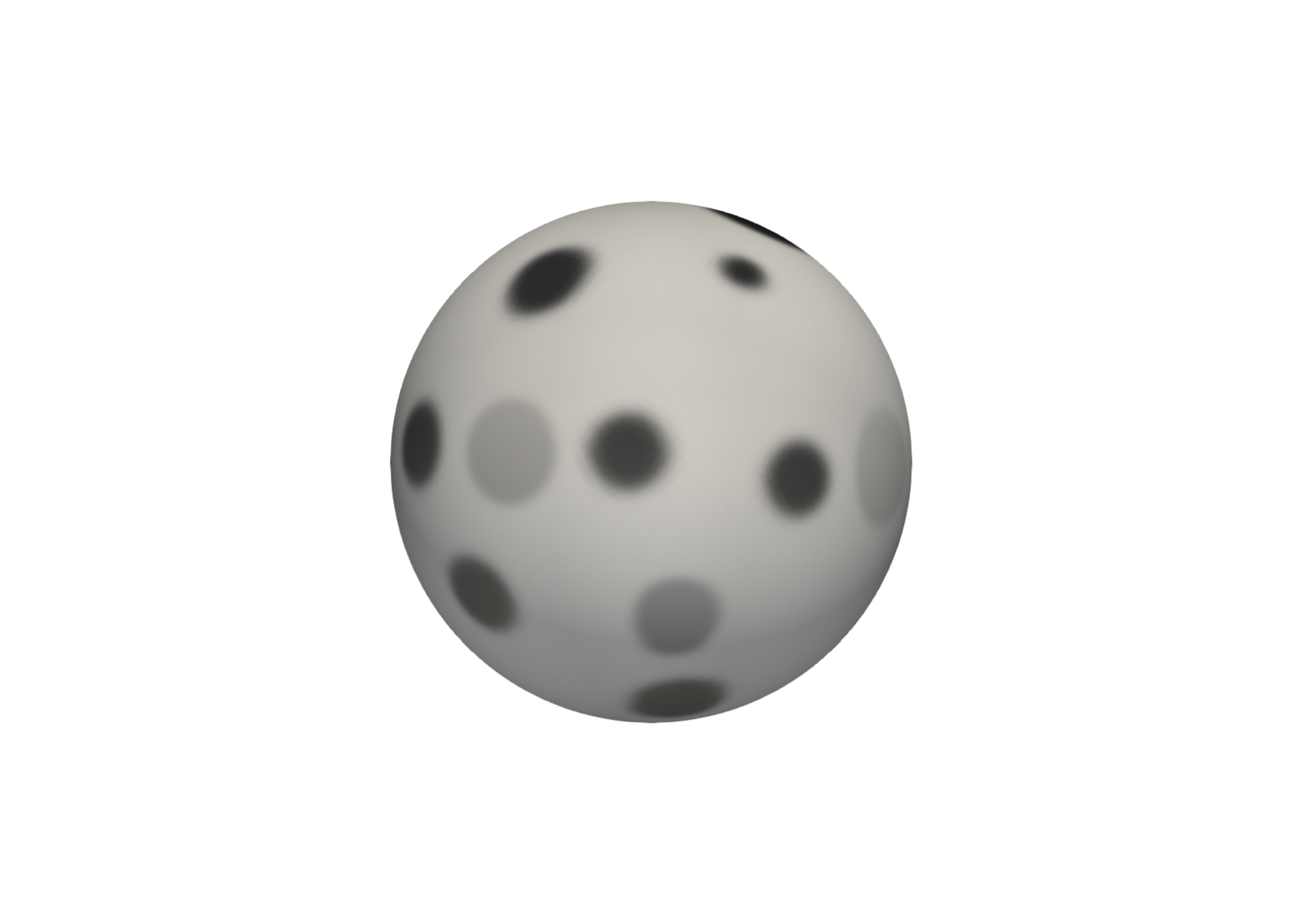}
\end{overpic}
\begin{overpic}[width=.2\textwidth,grid=false]{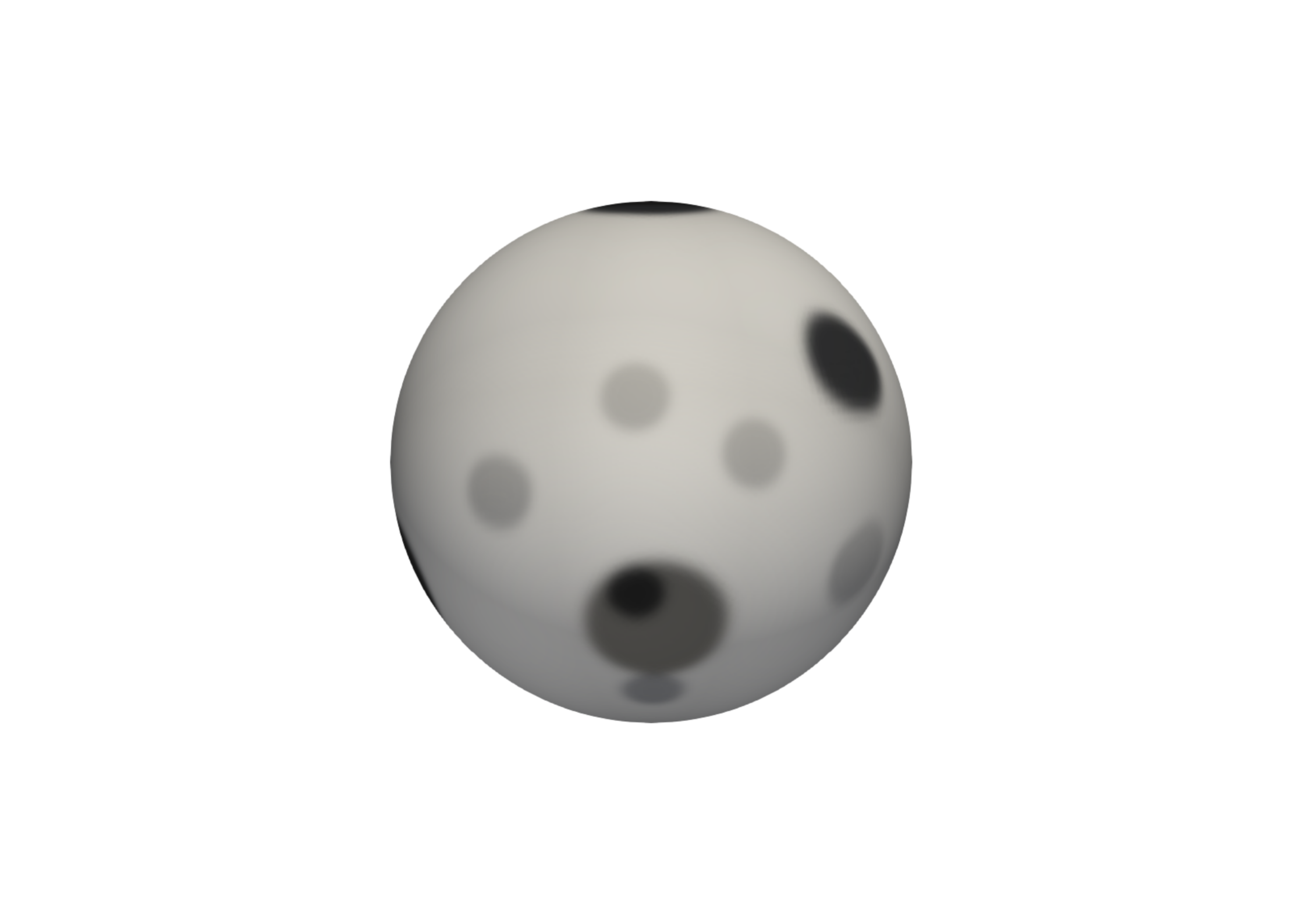}
\end{overpic}
\begin{overpic}[width=.2\textwidth,grid=false]{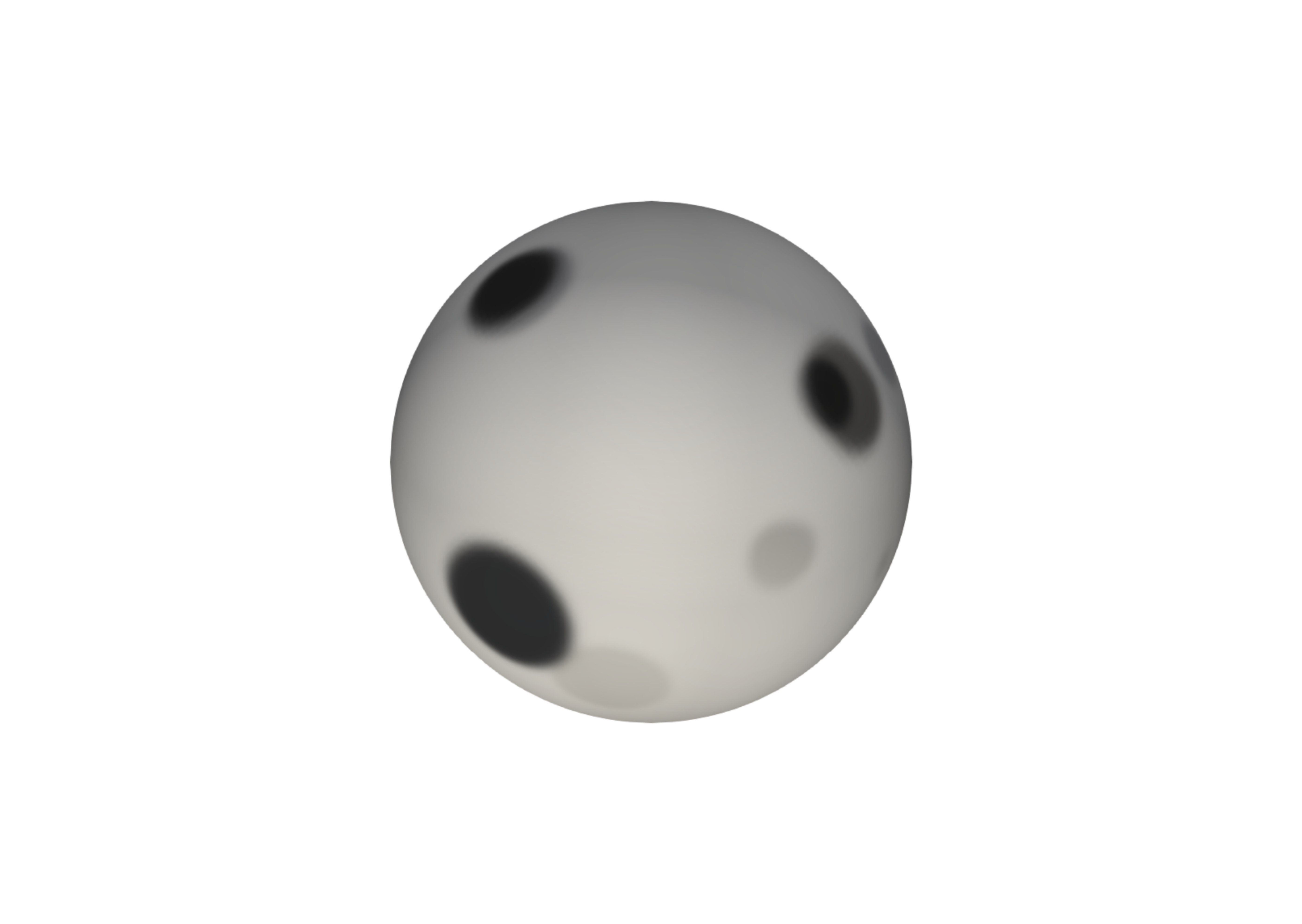}
\end{overpic}
\begin{overpic}[width=.2\textwidth,grid=false]{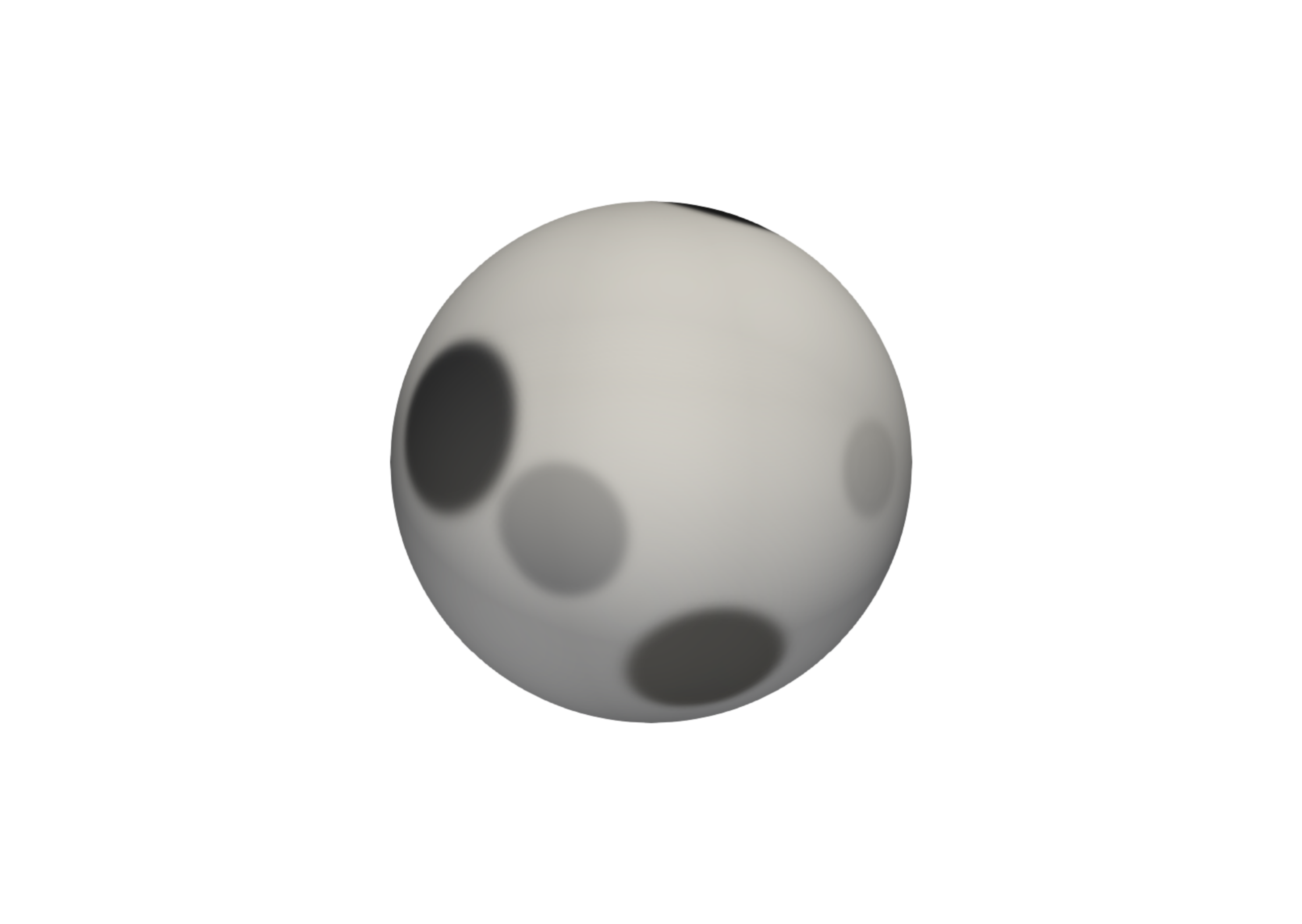}
\end{overpic}
}
\end{center}
\caption{Qualitative comparison for 3:1:20\%: epi-fluorescence microscopy images (with black background)
and numerical results (with white background) at eight different times in time interval $[156, 800]$ s. Click any picture above to run the full animation of a representative simulation.}\label{fig:qualitative_10}
\end{figure}

\begin{figure}
\begin{center}
\href{https://youtu.be/z2N9DmTGggQ}{
\begin{overpic}[width=.2\textwidth,grid=false]{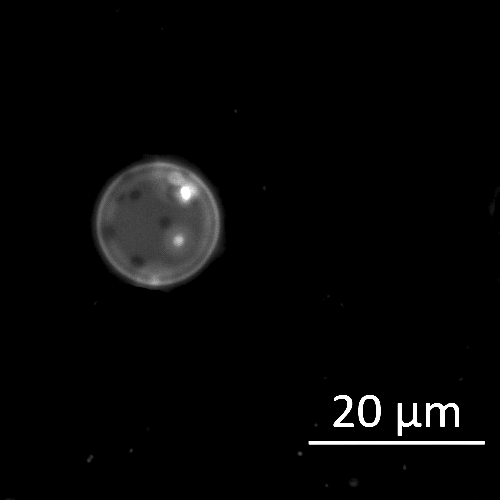}
\put(37,102){\small{$t = 124$}}
\end{overpic}
\begin{overpic}[width=.2\textwidth,grid=false]{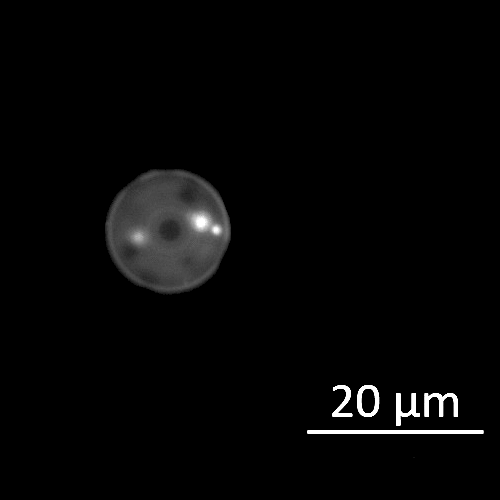}
\put(37,102){\small{$t = 148$}}
\end{overpic}
\begin{overpic}[width=.2\textwidth,grid=false]{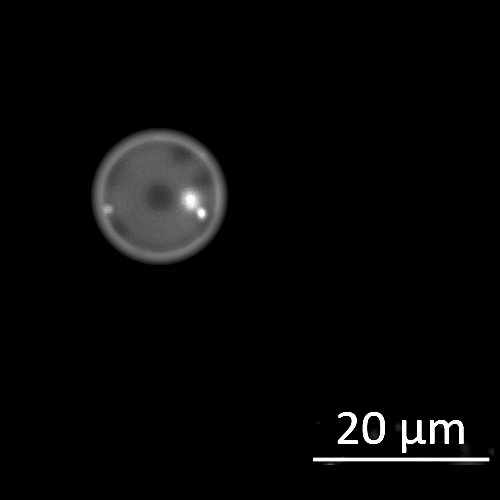}
\put(37,102){\small{$t = 172$}}
\end{overpic}
\begin{overpic}[width=.2\textwidth,grid=false]{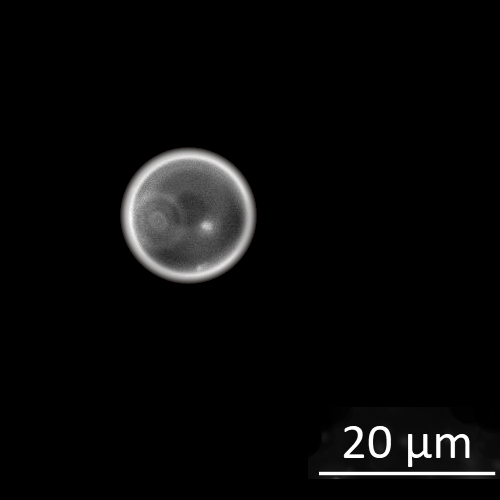}
\put(37,102){\small{$t = 294$}}
\end{overpic}
\\
%\vskip .5cm
\begin{overpic}[width=.17\textwidth,grid=false]{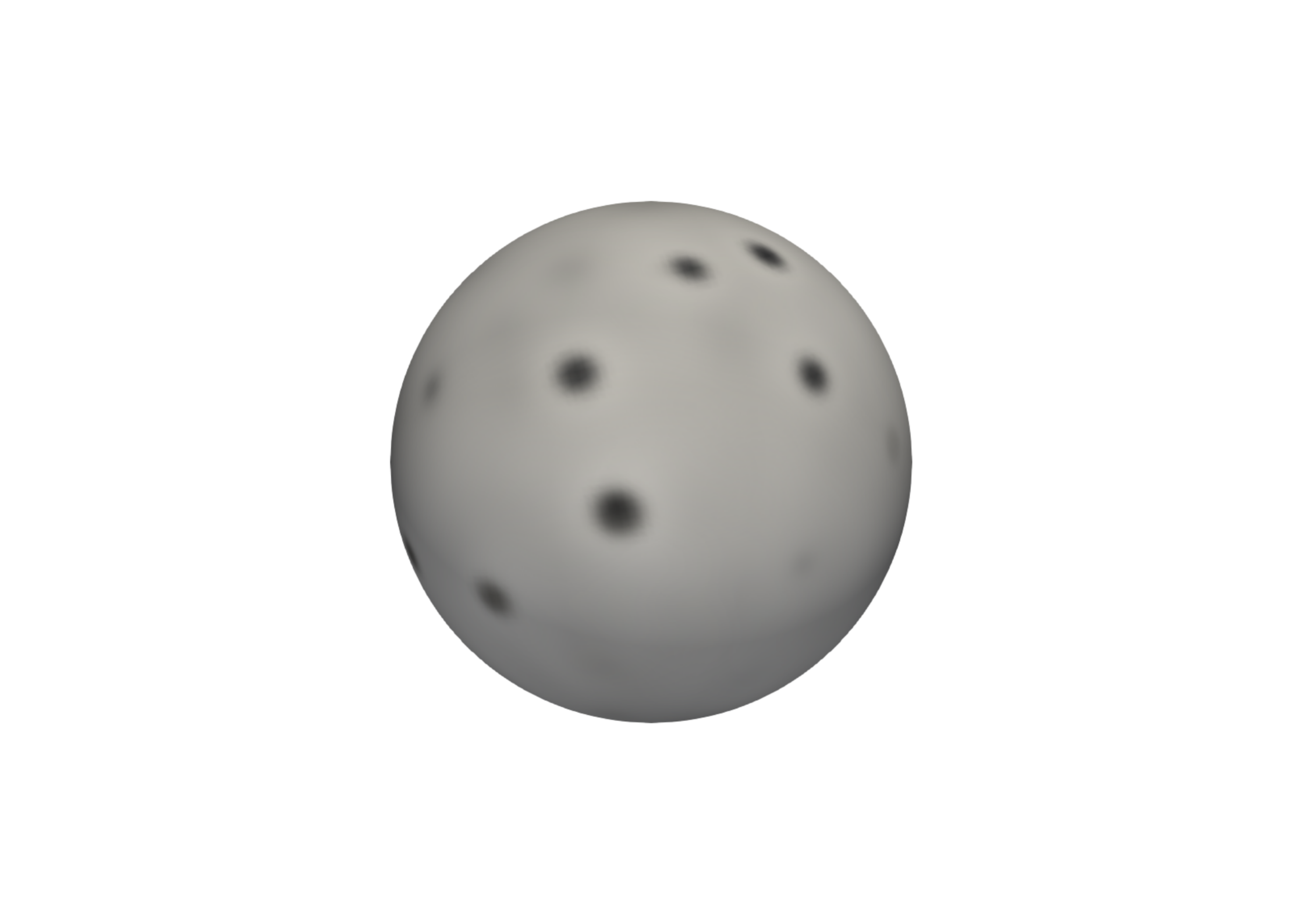}
\end{overpic}
\quad
\begin{overpic}[width=.17\textwidth,grid=false]{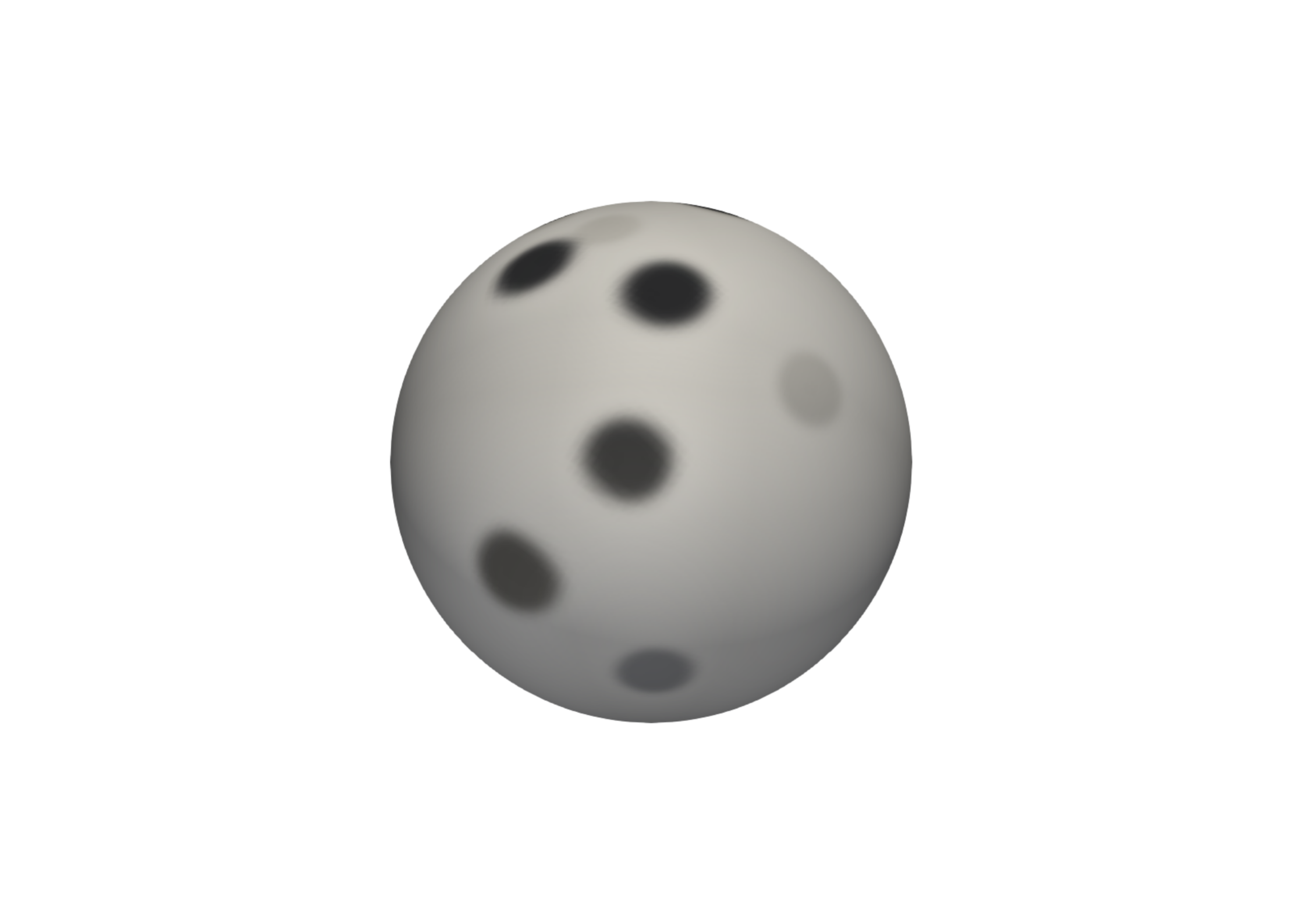}
\end{overpic}
\quad
\begin{overpic}[width=.17\textwidth,grid=false]{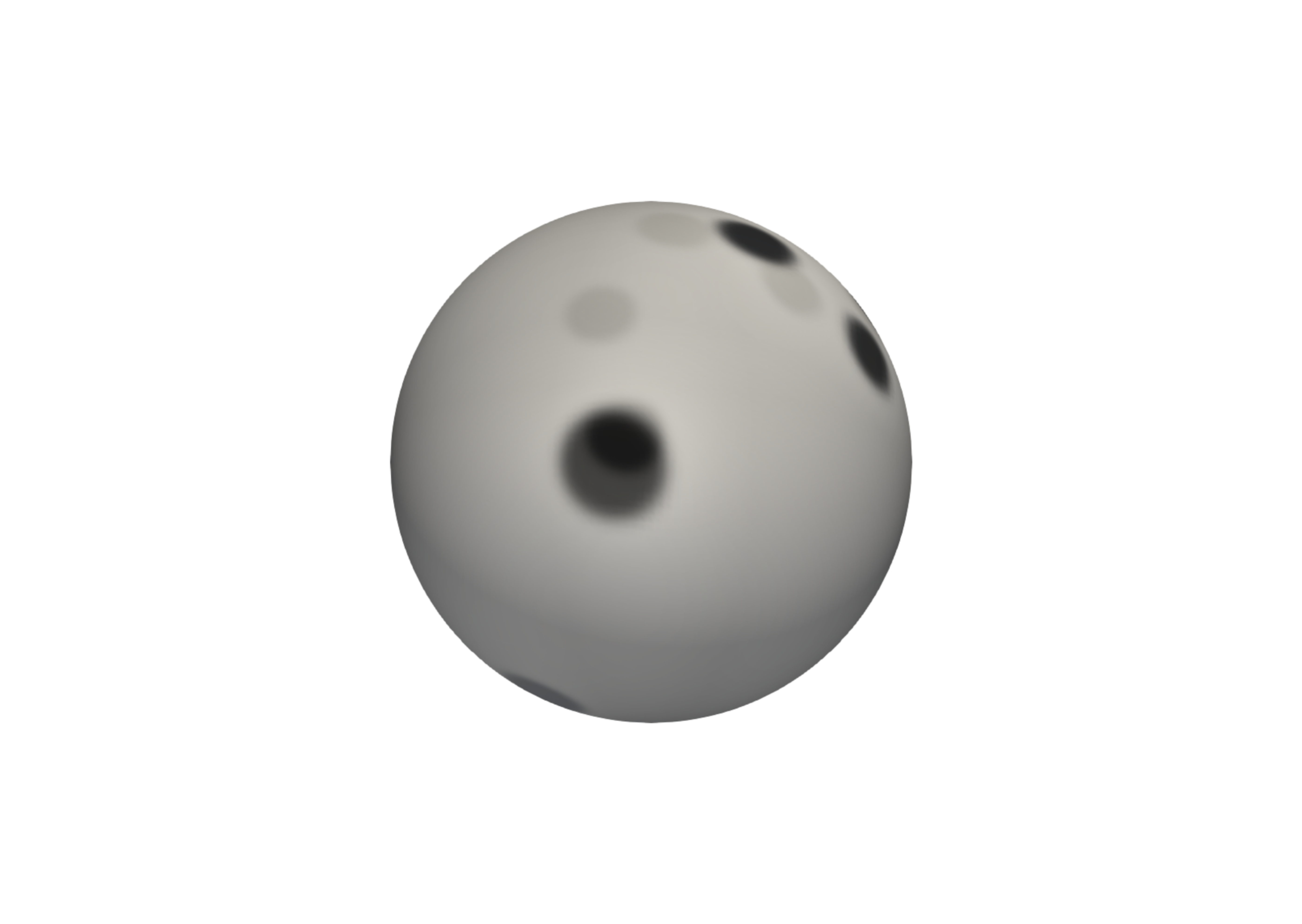}
\end{overpic}
\quad
\begin{overpic}[width=.17\textwidth,grid=false]{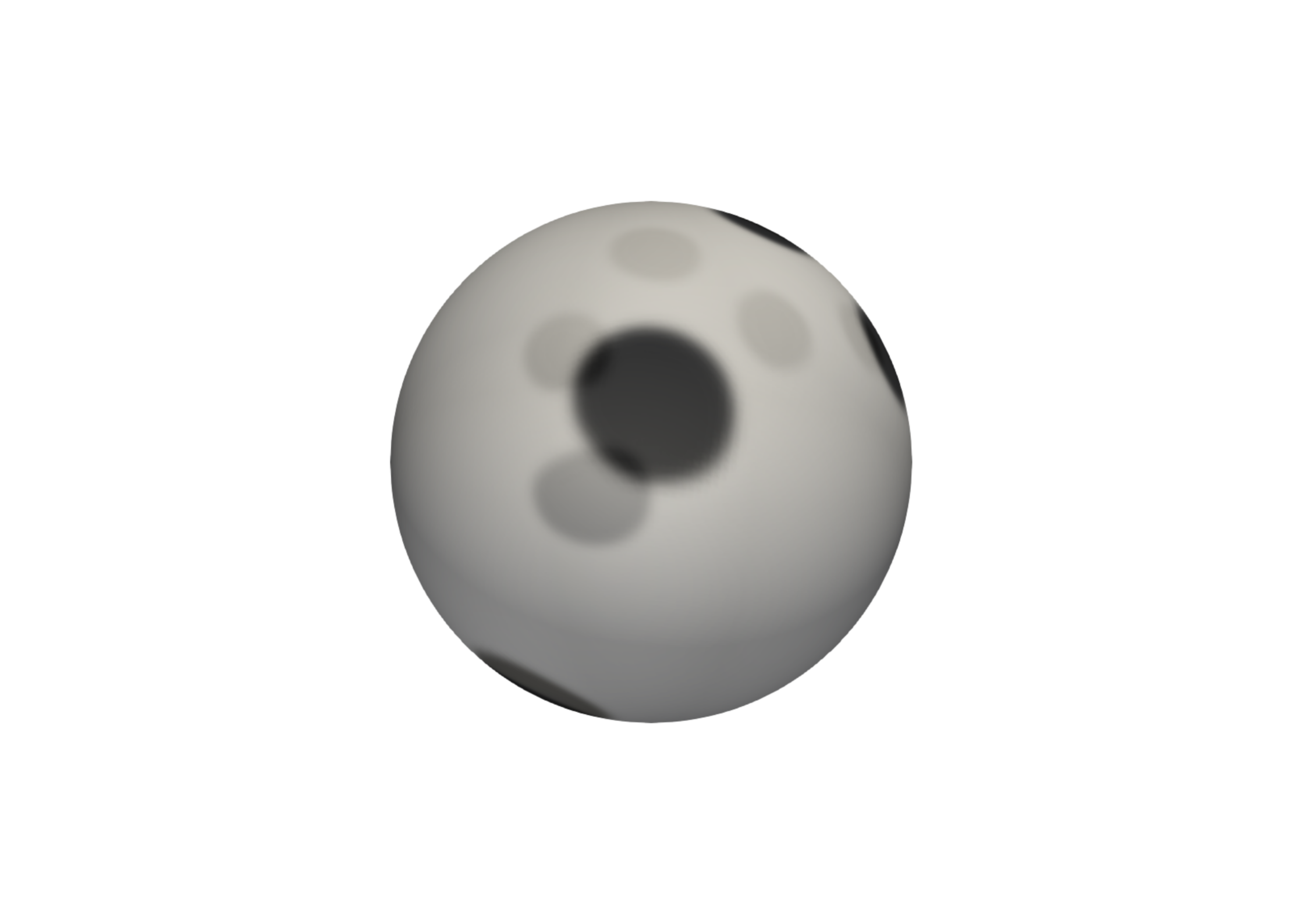}
\end{overpic}
}
\end{center}
\caption{Qualitative comparison for 2:1:20\%: epi-fluorescence microscopy images (with black background)
and numerical results (with white background) at four different times in time interval $[124, 294]$ s.
Click any picture above to run the full animation of a representative simulation.}\label{fig:qualitative_16}
\end{figure}

We conclude this section by presenting a qualitative comparison between images acquired with
epi-fluorescence microscopy and images obtained from post-processing the numerical results.
Fig.~\ref{fig:qualitative_10} (resp., Fig.~\ref{fig:qualitative_16}) presents such comparison for
composition 3:1:20\% (resp., 2:1:20\%).
Notice that the microscopy images in Fig.~\ref{fig:qualitative_10} and \ref{fig:qualitative_16} refer to different
sets of GUVs than those used for the quantitative analysis in Fig.~\ref{fig:raft_area}-\ref{fig:Raft}
because confocal microscopy (needed for the measurement) and epi-fluorescence microscopy cannot be used
simultaneously. Epi-fluorescence microscopy images could not be used for a quantitative
comparison with the numerical simulations because they provide only a two-dimensional picture
of the liposome. In post-processing the numerical results, we reduced the level of opacity
of the sphere representing the liposome to be able to see the rafts both in the front and in the back.
The rafts in the front are dark and should be compared with the rafts in the microscopy images,
while the rafts in the back are a lighter shade of gray.
%\anna{Do we want to comment on the bright artifacts in Fig.~\ref{fig:qualitative_16}?}
Overall, from Fig.~\ref{fig:qualitative_10} and \ref{fig:qualitative_16} we see an excellent
qualitative agreement between experiments and simulations.

%\begin{description}
%\item[Word] Definition
%\item[Concept] Explanation
%item[Idea] Text
%\end{description}

\section*{Conclusion}
This paper presents an experimental and computational study on the evolution of lipid rafts in
multicomponent membranes. Focusing on the ternary membrane composition DOPC:DPPC: Chol 
with well-known phase behavior, we studied domain formation on giant liposomes 
of two different molar ratios using fluorescence microscopy. 
Using state-of-the-art numerical techniques, we applied a continuum phase-field 
model to simulate domain formation on these liposomes. 
%We focused on two compositions: DOPC:DPPC with a 2:1 molar ratio
%with 20\% Chol and DOPC:DPPC with a 3:1 molar ratio with 20\% Chol.
%Our computational approach is based on a continuum model for membrane phase separation
%and uses state-of-the-art numerical techniques that became available only in recent years.
The numerical and experimental results are compared in terms of raft area fraction,
total raft perimeter over time and total number of rafts over time for both compositions
under consideration. Overall, excellent agreement is found.
To the best of our knowledge, this the first quantitative validation of a continuum based model
against experimental data.
These results show that this continuum model can provide accurate and quantitative 
prediction of lipid phase separation in membranes.

\section*{Acknowledgments}
The support of the University of Houston's Bridge Funds Program is gratefully acknowledged.
This work was also partially supported by US National Science Foundation (NSF) through grant DMS-1953535.
M.O.~acknowledges the support from NSF through DMS-2011444. 
S.M.~acknowledges the support from NSF through DMR-1753328. 
The authors are grateful to Sergei Mukhin, Timur Galimzyanov,
Boris Kheyfets and Vassiliy Lubchenko for insightful discussions on thermodynamic properties of lipid bi-layers. 

% Uncomment if using bibtex (default)
\bibliographystyle{plain}
\bibliography{literatur}

\end{document}